\documentclass[conference]{IEEEtran}
\IEEEoverridecommandlockouts
\usepackage{times}                     
\usepackage{float}
\usepackage{subfigure}
\usepackage{amsmath}
\usepackage{algorithm}
\usepackage{algorithmicx}
\usepackage{algpseudocode} 
\usepackage{tabu}                      
\usepackage{booktabs}                  
\usepackage{lipsum}                    
\usepackage{mwe} 
\usepackage{array}

\def\BibTeX{{\rm B\kern-.05em{\sc i\kern-.025em b}\kern-.08em
    T\kern-.1667em\lower.7ex\hbox{E}\kern-.125emX}}
\begin{document}

\title{NeRFlex: Resource-aware Real-time High-quality Rendering of
Complex Scenes on Mobile Devices\\
\thanks{\IEEEauthorrefmark{1}Corresponding author.}

}
\author{\IEEEauthorblockN{Zhe Wang and Yifei Zhu\IEEEauthorrefmark{1}}
\IEEEauthorblockA{\textit{UM-SJTU Joint Institute} \\
\textit{Shanghai Jiao Tong University, Shanghai, China}\\
123369423@sjtu.edu.cn, yifei.zhu@sjtu.edu.cn}
}

\maketitle
\begin{abstract}
Neural Radiance Fields (NeRF) is a cutting-edge neural network-based technique for novel view synthesis in 3D reconstruction. However, its significant computational demands pose challenges for deployment on mobile devices. While mesh-based NeRF solutions have shown potential in achieving real-time rendering on mobile platforms, they often fail to deliver high-quality reconstructions when rendering practical complex scenes. Additionally, the non-negligible memory overhead caused by pre-computed intermediate results complicates their practical application. To overcome these challenges, we present NeRFlex, a resource-aware, high-resolution, real-time rendering framework for complex scenes on mobile devices. NeRFlex integrates mobile NeRF rendering with multi-NeRF representations that decompose a scene into multiple sub-scenes, each represented by an individual NeRF network. Crucially, NeRFlex considers both memory and computation constraints as first-class citizens and redesigns the reconstruction process accordingly. NeRFlex first designs a detail-oriented segmentation module to identify sub-scenes with high-frequency details. For each NeRF network, a lightweight profiler, built on domain knowledge, is used to accurately map configurations to visual quality and memory usage. Based on these insights and the resource constraints on mobile devices, NeRFlex presents a dynamic programming algorithm to efficiently determine configurations for all NeRF representations, despite the NP-hardness of the original decision problem. Extensive experiments on real-world datasets and mobile devices demonstrate that NeRFlex achieves real-time, high-quality rendering on commercial mobile devices.
\end{abstract}


\section{Introduction}
Reconstructing 3D scenes and objects with high accuracy from 2D RGB images presents a significant challenge in computer vision and graphics.
Achieving high-fidelity and fast reconstruction of realistic scenes on mobile devices has important applications in robotics, gaming, and AR/VR\cite{singandhupe2019review}\cite{cheng2022review}\cite{lee2023farfetchfusion}. In these interactive applications, delays or quality degradation can severely impact user experience. 
In recent years, Neural Radiance Fields(NeRF) has gained widespread attention as a novel implicit rendering technique for 3D reconstruction. Compared to traditional approaches like point clouds or meshes, NeRF provides photo-realistic rendering while significantly reducing storage requirements. NeRF trains a neural network that takes the spatial coordinates and viewing directions as input and then outputs the color and density of each pixel by inferring all the sampled points along the ray. Consequently, NeRF algorithms must execute the neural network hundreds of millions of times for a single view, leading to considerable computational and time costs \cite{mildenhall2021nerf}.

To fully unleash the potential of NeRF techniques, existing studies pre-compute and store intermediate results of the NeRF network, typically represented as voxels or meshes, to facilitate rendering. 
This approach avoids inferring the entire network during rendering, thereby accelerating the rendering process and reducing computational costs\cite{hedman2021baking}\cite{chen2022mobilenerf}\cite{tang2023delicate}. Other approaches focus on modifying the network itself, such as simplifying the object into sparse voxel grids \cite{liu2020neural} or reducing the number of sample points per ray and accelerating the access to each data point \cite{muller2022instant}. To represent large-scale, realistic scenes, researchers propose dividing the complex scene into smaller regions and representing each with an individual NeRF neural network. The outputs of these NeRFs are then integrated to generate the final rendering\cite{tancik2022block}\cite{liu2023real}.

While the aforementioned methods can improve the rendering speed of NeRF or representation capability to some extent, a practical solution that delivers both high-quality and real-time complex scene rendering on mobile devices is still lacking. This is a challenging task for several reasons.  First, complex scene rendering demands higher quality. Unlike the synthetic single-object scenes that most studies focus on, real-world scenarios consist of multiple objects with distinct semantic meanings. Treating an entire scene as a single object using one neural network reduces rendering quality because each training sample must encompass all objects. This can cause complex objects to occupy only a small number of pixels in the training frame, significantly degrading the NeRF network's ability to reconstruct those regions. 
Second, representing complex scenes imposes significant rendering costs on mobile devices in terms of both time and memory. Complex scenes have fewer empty pixels, and rendering each pixel requires hundreds of inferences through the entire NeRF network. While existing multi-NeRF solutions may help with scene representation, they overlook the additional resource demands introduced by multiple NeRF networks, making them impractical for mobile deployment.

To fill in this gap, we propose \textit{NeRFlex}, a system designed for real-time, high-quality rendering of complex scenes on mobile devices. Inspired by existing efforts in mesh-based real-time NeRF rendering and multi-nerf solutions for large-scale scene representation, NeRFlex adopts a divide-and-conquer strategy by decomposing a complex scene into multiple sub-scenes. It further adaptively adjusts the multi-modal NeRF data for each sub-scene to meet device constraints while optimizing rendering quality. This is non-trivial and necessitates solving the following two major challenges. 

The first challenge is determining which sub-scenes are worth representing with a network. Simply assigning a separate network to each semantically meaningful object in the scene can quickly exceed the computational and memory capacities of mobile devices, resulting in rendering failures. Therefore, it is crucial to devise an efficient strategy for decomposing the scene and identifying which objects or regions warrant individual mesh-based NeRF representations.

The second challenge is finding out the appropriate representations can be costly and computationally hard. 
For the mesh-based NeRF rendering algorithm, the saved multi-modal representation data significantly affects the memory and computation cost of rendering these networks. While finer representations provide higher quality, they also demand more mobile resources \cite{chen2022mobilenerf}. Conducting a brute-force search for all possible configurations of all objects to find the optimal one introduces prohibitively high search costs, as this problem is essentially NP-hard.

To overcome these challenges, NeRFlex proposes a detail-based scene segmentation module to identify worthwhile sub-scenes based on domain knowledge. It further enhances these scenes through interpolation, enabling the network to learn them more effectively. NeRFlex then builds a lightweight profiler that accurately captures the relationship between various configurations and resource consumption or rendering quality. Using this information, and considering device resource constraints, NeRFlex employs a dynamic programming algorithm to efficiently determine the optimal configuration for all NeRF networks within the exponential search space.  
The contributions of our work can be summarized as follows:
\begin{itemize}
    \item NeRFlex presents an efficient volumetric rendering solution for complex scene rendering on mobile devices by judiciously integrating divide-and-conquer principles into mesh-based rendering. 
    \item NeRFlex employs detail-oriented semantic segmentation to represent objects of interest with high-frequency details. It incorporates a lightweight profiler that accurately captures the relationship between configuration and performance metrics without requiring costly trial-and-error processes. Our domain knowledge in computer graphics informs efficient and novel system designs for importance evaluation and constructing profiling models.
    \item NeRFlex designs an efficient dynamic programming algorithm to adaptively adjust configuration parameters for each segmented sub-scene despite the NP-hardness of the original problem. 
    \item Extensive evaluations on mobile devices demonstrate that NeRFlex successfully achieves smooth complex scene rendering on mobile devices while maximizing rendering quality.
\end{itemize}

\section{Background and Related Work}

\subsection{NeRF Rendering Acceleration}
While NeRF delivers impressive photo-realistic 3D reconstruction quality, its rendering process is notably time-intensive, requiring millions of neural network evaluations per pixel. To achieve real-time rendering speeds, and even to free the NeRF rendering from dependence on high-performance hardware, researchers have explored two primary strategies. The first strategy involves simplifying the scene or object, allowing the network to focus solely on learning the sparse voxel representation or features of the target \cite{liu2020neural}\cite{cao2023real}\cite{sun2022direct}\cite{li2022vox}. Furthermore, utilizing various data structures, such as hash tables, can significantly reduce processing times \cite{muller2022instant}\cite{deng2023compressing}. The second strategy focuses on pre-computing partial results from the original network, which are then stored as features of the scene or objects. This approach is commonly referred to as the baked NeRF method \cite{hedman2021baking}\cite{chen2022mobilenerf}\cite{tang2023delicate}\cite{gao2210nerf}. Essentially, this approach substitutes the original computational demands of the neural network with a minimal MLP, significantly reducing inference time.  However, these works primarily focus on the NeRF rendering or training acceleration, leaving the fidelity issue in representing complex scenes untouched. In contrast, our system not only aims to reduce the computational demands on mobile devices but also ensures high-quality rendering for complex scenes.
From the system perspective, some preliminary results have identified the effects of different configurations on system metrics\cite{chen2024nerfhub}\cite{wang2024towards}. NeRFhub \cite{chen2024nerfhub}, one of the latest efforts,
determines the Pareto hyper-parameters for the multi-layer perception network pruning used in NeRF, along with feature image and 3D data in the post-processing stage. In contrast, rather than applying optimizations after the feature images and 3D data are generated, NeRFlex directly optimizes the training and generation process itself, ensuring that the outputted data is resource-efficient while maintaining high rendering quality.

\subsection{Complex Scene NeRF Rendering}
For complex scenes, particularly large-scale environments, research efforts are typically categorized into several distinct approaches: spatial scene division \cite{turki2022mega}\cite{tancik2022block}\cite{liu2023real} and integrated scene representation \cite{barron2022mip}\cite{zhang2024efficient}\cite{xiangli2022bungeenerf}. Spatial scene division involves segmenting the scene into numerous cells, with each cell being trained by its own dedicated network, which significantly enhances the model's ability to capture details across extensive areas. In contrast, some researchers strive to consolidate the entire scene into a single NeRF network to reduce computational requirements while maintaining image quality. 
To further improve the rendering quality, researchers have achieved promising results by altering network structures \cite{kerbl20233d}\cite{martin2021nerf}\cite{barron2022mip} and introducing new mechanisms \cite{rakotosaona2024nerfmeshing}\cite{tang2023delicate}. However, directly applying these approaches often increases the computational load on top of the original network's requirements, challenging their deployments on mobile devices with limited computational capacity. 
In the context of mobile deployment for complex scene NeRF rendering, current efforts focus on transforming pre-trained NeRF variant networks into representations with low computational demands. 
Building on the concept of distillation, a lightweight convolutional model is employed to approximate the output of a well-trained, high-quality NeRF model. As a result, the convolutional model can be used for rendering directly, without relying on the computationally intensive NeRF model, thus meeting the constraints of mobile devices \cite{cao2023real}.
Similarly, the approach in \cite{rojas2023re} converts well-pre-trained NeRF representations into a more efficient form by extracting density information into a grid and factorizing color data into a light field. This approach avoids hundreds of MLP inferences by using matrix multiplication for rendering. However, these methods exhibit a fundamental limitation: their rendering performance often fails to match or exceed that of the pre-trained NeRF, as they do not improve or refine the initial full-scale NeRF model's quality. Consequently, the inherent shortcomings of the pre-trained NeRF are retained or even magnified.

In contrast, we aim to ensure high-quality rendering of complex scenes while significantly reducing the hardware requirements. This enables local rendering on devices with lower computational capabilities, making advanced rendering techniques more accessible and practical for a wider range of mobile platforms.

\section{System Design}
\begin{figure}[!tbp]
  \centering
  \includegraphics[width=\linewidth]{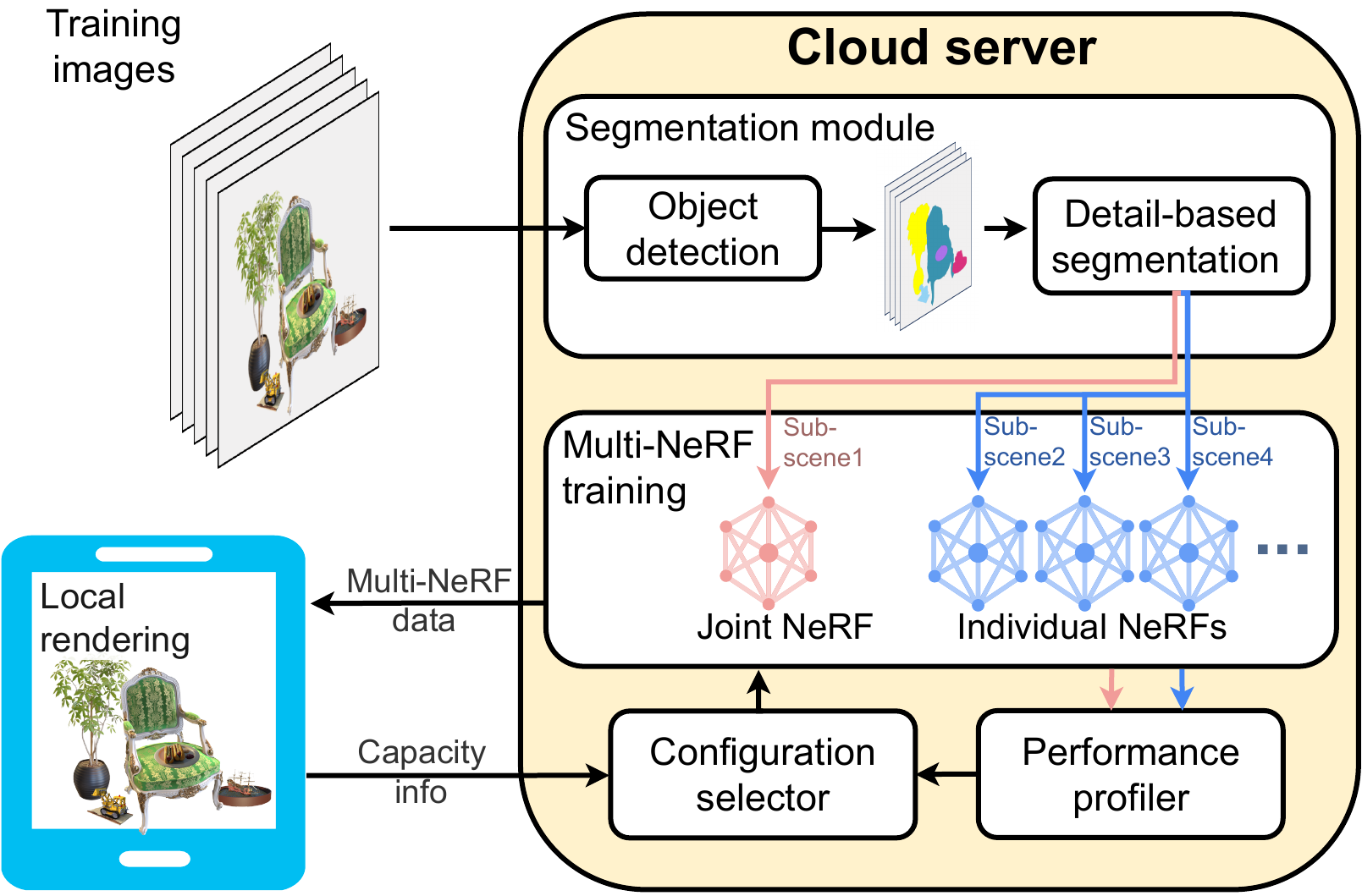}
  \caption{System overview of NeRFlex, an on-device real-time, high-quality, complex scene rendering system.
  }
  \label{pipeline}
\end{figure}

Figure \ref{pipeline} provides an overview of our system design. Instead of directly training the NeRF network with all these input images, NeRFlex introduces three modules to realize high-quality, real-time complex scene rendering on mobile devices. 
The training images are first processed by a segmentation module. It segments objects in the scene based on their complexity of detail, identifying those that merit individual representation by a separate NeRF network. After that, lightweight profiling models are generated for each segmented object to quantify the impact of different configurations on memory and quality. Based on this information and the collected memory capacity information, the configuration selector then identifies the configuration pair for the representation data with the highest rendering quality under the size limitation. Then, these configuration decisions are used to guide the parallel training for the NeRF networks. The trained NeRF networks and their corresponding intermediate mesh data and feature maps are then sent to the mobile devices for local rendering.  

\subsection{Detail-based Segmentation}

NeRFlex adopts a divide-and-conquer approach, representing the whole scene with multiple NeRFs. This design abstracts away network details and can be directly applied to all the NeRF variants without any modifications or requirements on the network itself. 

The starting segmentation module aims to identify objects or sub-scenes in the scene that warrant individual representation by a single NeRF. It then automatically separates them from the training photo set to create a dedicated training set for training the corresponding NeRF. In this way, the training images for each NeRF only need to contain the single object's content. Compared with the original training images, we keep the same image size (the same number of pixels) but retain and enlarge the target object, reducing the frequency of details the network needs to learn.

\begin{figure}[!tbp] 
  \centering
  \includegraphics[width=\linewidth]{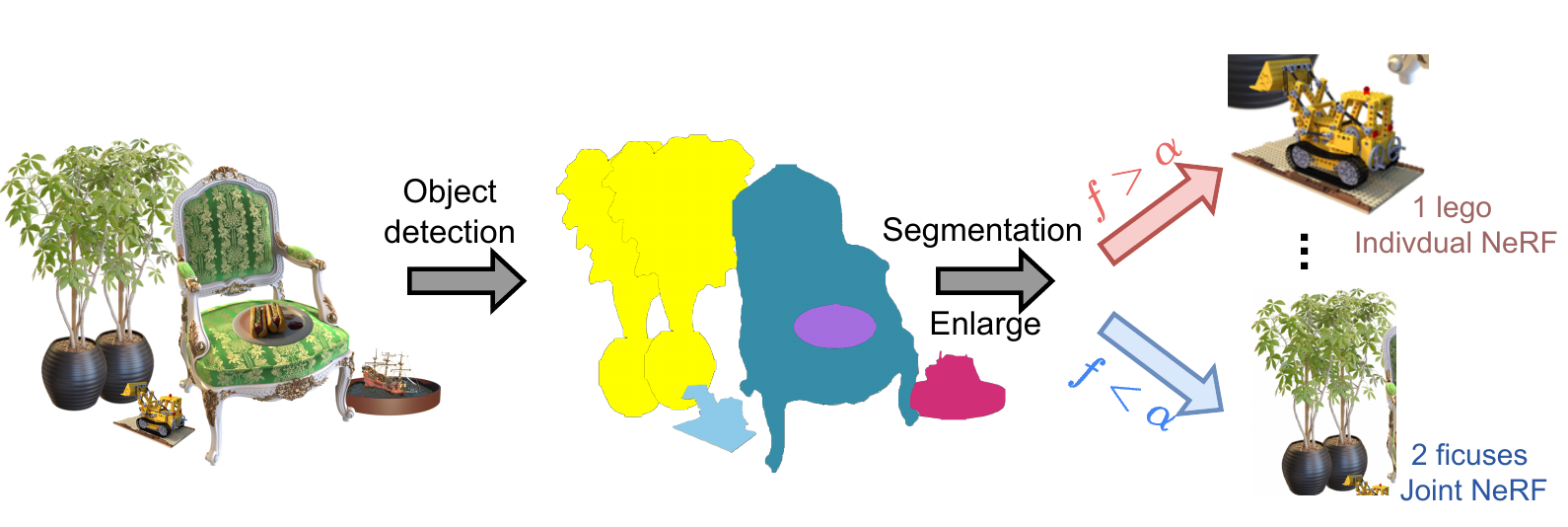}
  \caption{Working pipeline of the segmentation module. $f$ refers to the maximum frequency of the object, while $\alpha$ refers to the threshold.}
  \label{seg}
\end{figure}

The module can be divided into two sub-modules, object detection and detail-based segmentation. Figure \ref{seg} illustrates the module's working pipeline. Object detection is first applied to all the original training images to detect objects in these images that can be segmented and generate a corresponding mask to cover all the pixels they occupy, preparing for subsequent segmentation.
For each detected object in each image, the detail frequency of the object within that image is also calculated and recorded. 
Then we use the maximum frequency recorded for each object to determine whether it merits representation by a separate network. Compared to using average frequency as a segmentation metric, the recorded maximum frequency better reflects the importance of an object to the user's viewing experience. 

After determining the maximum frequency for each corresponding object, a threshold frequency value is established to decide which objects warrant individual NeRF representations. If an object's maximum frequency exceeds this threshold, it is assigned a dedicated NeRF. Otherwise, it is represented collectively with other objects whose maximum frequencies fall below the threshold, utilizing a single NeRF network. This threshold can be adjusted by users to better suit their specific needs.

After identifying which objects in the scene merit separate NeRF representation based on the original training images, we extract these objects from each image based on the mask, using its outermost pixels as boundaries. We then appropriately scale these segmented parts using interpolation scaling to create a new image for further training of the corresponding NeRF network. This process effectively reduces the frequency of the detailed regions or faces of the objects that need to be learned, leading to improved rendering quality.

This design is based on two observations using domain knowledge. First, single NeRF has a weak learning ability for high-frequency details and it results in lower rendering quality in those regions \cite{mildenhall2021nerf}. So objects with high frequency are more crucial for NeRF rendering. Second, when viewing an object, users often focus on the more complex side or high-frequency regions or faces of the objects \cite{zhang2017novel}\cite{hou2007saliency}. Therefore, using the maximum frequency from different viewpoints (images) as an indicator can more accurately reflect the user's focus on an object and its importance to the viewing experience.

\subsection{Lightweight Profiling}
\begin{figure}[!tbp]
    \centering
    \subfigure[The quality fitted curve validation with various mesh granularity levels on patch size of 17.]{\includegraphics[height=3.7cm,width=0.49\linewidth]{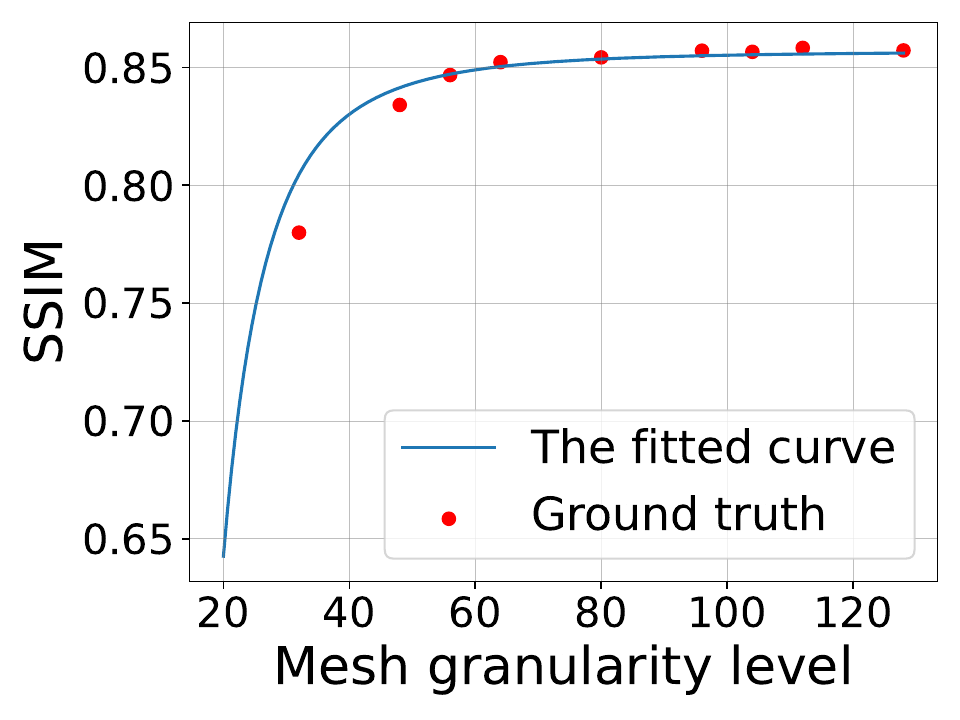}\label{gran_p17}}
    \subfigure[The size fitted curve validation with various mesh granularity levels on patch size of 17.]
    {\includegraphics[height=3.7cm,width=0.49\linewidth]{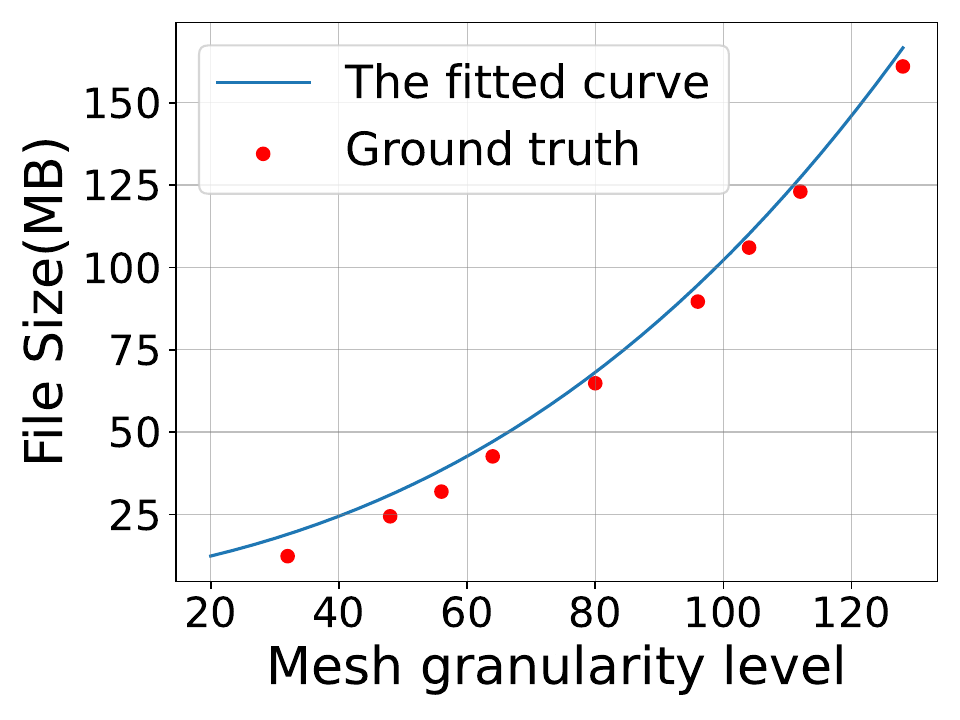}\label{size_p17}}
    \subfigure[The quality fitted curve validation with various patch sizes on mesh granularity level of 80.]{\includegraphics[height=3.7cm,width=0.49\linewidth]{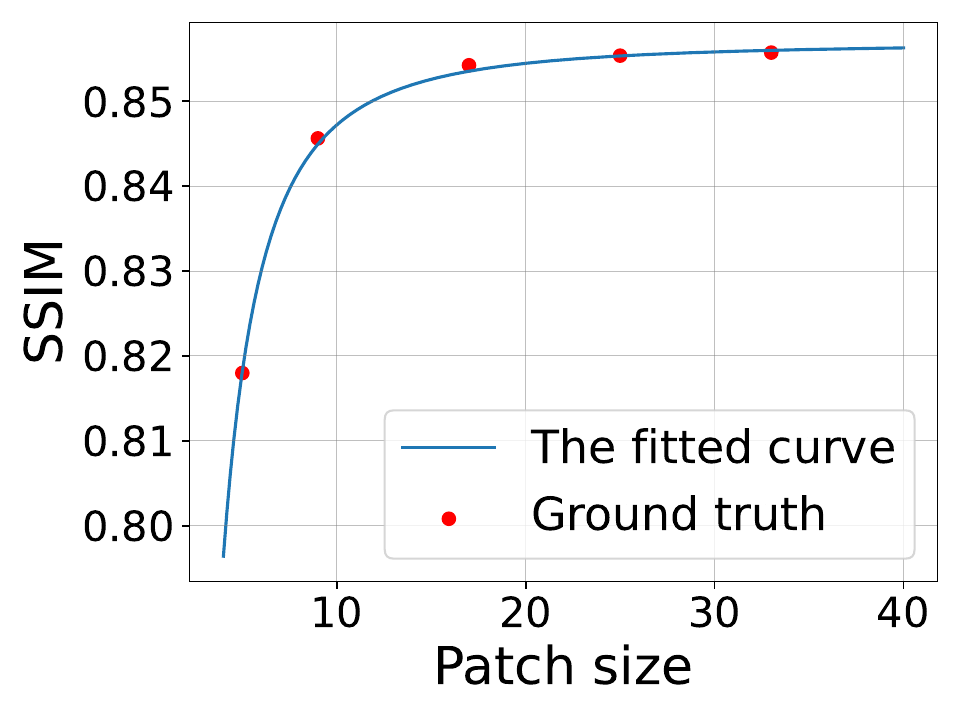}\label{gran_g80}}
    \subfigure[The size fitted curve validation with various patch sizes on mesh granularity level of 80.]
    {\includegraphics[height=3.7cm,width=0.49\linewidth]{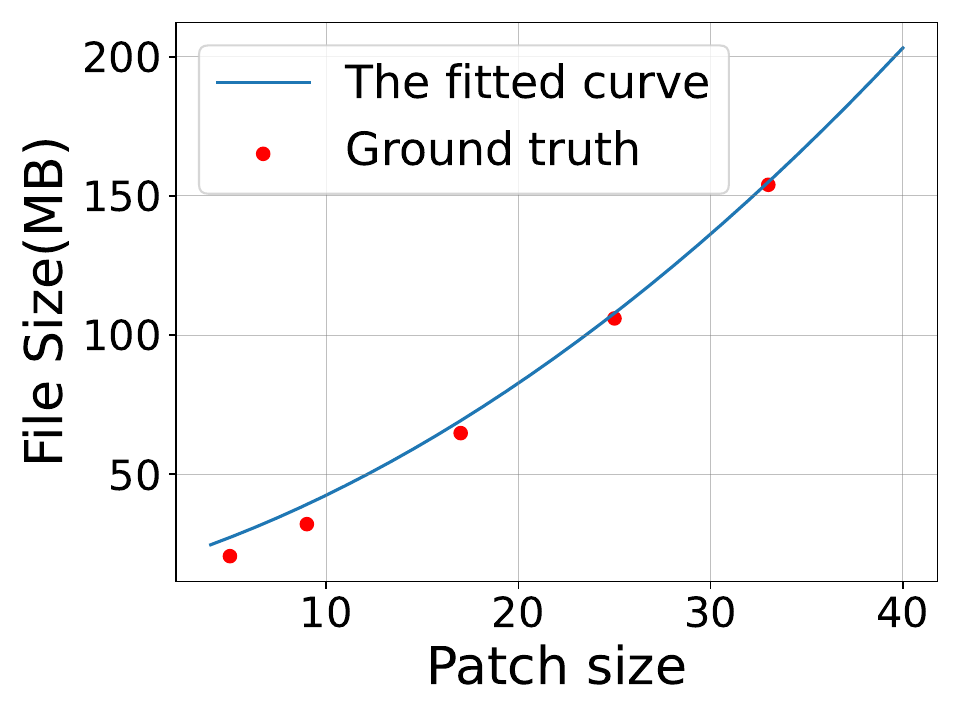}\label{size_g80}}
    \caption{Emperical evaluation of our profiling models}
    \label{model_curve}
\end{figure}

Only mesh-based NeRF networks and their auxiliary data with proper sizes can fit into mobile devices. Jointly they affect the rendering quality as well. It is extremely time-consuming and resource-consuming to go through all possible representations under different configurations and identify the right one. According to experimental data, training a NeRF network with the configuration provided in the original paper \cite{chen2022mobilenerf} often takes several hours on four RTX 3090 GPUs. Moreover, the higher the configuration requirements for the baked data, the longer the training time required.

Therefore, NeRFlex designs a profiling module that predicts the corresponding object's representation data size as well as the rendering quality under different configurations. It can directly assess the value and loss provided by a configuration pair without the need to train the entire neural network, avoiding significant time and resource costs. 
We use the SSIM (Structural Similarity Index Measure) metric as the quality metric, denoted as $Q$. This metric evaluates luminance, contrast, and structure between images \cite{wang2004image}, offering a closer alignment with human visual perception compared to PSNR metric, which only assesses pixel color differences.  We use the total data size to quantify the load for memory, denoted as $S$.

For our focused mesh-assisted real-time NeRF solutions, e.g., mobile-NeRF \cite{chen2022mobilenerf} and NeRF2Mesh \cite{tang2023delicate}, the objects are always first converted to a voxel grid representation. A higher grid density typically results in more accurate mesh reconstruction, as allocating more meshes enhances the overall reconstruction quality.
In the meantime, the texture image is also generated. It is used to color all the quad faces of the meshes to form the final appearance. For each quad face, they allocate $p\times p$ pixels for its final appearance texture. The impact of texture size on rendering quality is similar to that of voxel grid density. Considering that the voxel grid representation is three-dimensional, we set the controlling knob in geometry mesh as the number of voxel grid allocated for each axis, denoted as $g$; we set the one-dimensional size, $p$ of texture image as another controlling knob in the texture image. We exclude the configuration knob for the MLP, as its size is extremely small, around only a few KB. Additionally, we found that MLP quantization has poor compatibility with existing commercial rendering engines, consistent with previous findings \cite{wang2024towards}. Therefore, we did not consider adjusting the MLP in our approach.

Instead of resorting to complex black-box models, e.g., neural networks, to build this profiling model, we build a white-box profiling model based on our knowledge in the rendering. 
During training, NeRF divides the entire rendering space into $g^3$ voxels, and meshes are subsequently formed based on neighboring voxels. As the granularity increases, more voxels are generated for mesh construction, resulting in a more detailed mesh model. Texture patches are assigned to these meshes, with each mesh receiving a patch size of $p^2$. The rendering quality improves as the patch size increases, as larger patches can store more texture encoding information.
The baked data size is primarily composed of the geometric mesh and the corresponding texture images. The size of the geometric mesh is determined by the number of meshes, which in turn depends on the number of voxels ($g^3$) the scene is divided into. The size of the texture images is influenced by both the number of meshes and the number of texture pixels ($p^2$) allocated to each quad face.
Thus, we express both the reconstruction quality and the baked data size as polynomials of $g$ and $p$.
\begin{equation}
    \left\{
        \begin{aligned}
        &S=f_s({\theta}) =  f_s(g, p) = \frac{-k}{(g + a)^3 * (p + b)^2} + m\\
        &Q=f_q({\theta}) =  f_q(g, p) = k^{'} * (g + a^{'})^3 * (p+b^{'})^2\\
        \end{aligned}
    \right. 
\label{pre_model_com}
\end{equation}
Due to the simple form of the profiling models, except for $g$ and $p$, all the other parameters can be easily determined through curve fitting. To further minimize the number of sampling points for curve fitting, we design a variable step-size searching strategy within NeRF's configuration space. Specifically, for selecting the $g$ values of the sample points, the step size is $2*g'$, where $g'$ represents the value of the previous sample point.
For each $g$ value, we select the maximum, minimum, and midpoint values of the patch size range as three distinct $p$ values, pairing them with $g$ to generate different sample points.

We randomly select an object from the 3D object dataset provided by \cite{mildenhall2021nerf} to validate the performance of our profiling models. Figure \ref{model_curve} presents the accuracy of our prediction model on a typical mesh-assisted NeRF algorithm. To assess the accuracy of the model for each controlling knob, we held the other knob constant. Figure \ref{gran_p17} and \ref{size_p17} display the fitting results for the mesh control knob against the ground truth in terms of both quality and size, with the texture control knob fixed at 17. Figure \ref{gran_g80} and \ref{size_g80} show the fitting results for the texture control knob under a fixed mesh control knob of 80.
To further comprehensively validate the performance of our proposed profiler, we conduct error analysis experiments on four different objects with 45 configuration pairs. The experiment results demonstrate that the average error in quality prediction is 0.0065 with a standard deviation of 0.0088 for SSIM, and the average error in size prediction is 3.3410 with a standard deviation of 2.7345. These results clearly demonstrate that our profiling models accurately capture the relationship between configuration pairs and rendering quality, as well as the resulting file sizes.

\subsection{DP-based Configuration Selector}
The configuration selector module aims to figure out a configuration pair, which can achieve the highest rendering quality within a given space constraint, for the mesh-assisted NeRF rendering.

After segmenting high-frequency detail objects and generating their respective profiles, the next task is to select the suitable configuration pair for the corresponding mesh-assisted NeRF. Existing research on configuration selection for baked NeRFs typically focuses on single-object representation \cite{chen2024nerfhub}. Additionally, the configuration knobs studied are often related to the compression algorithms of the baked data or the number of parameters in the NeRF network, which differ significantly from our design. In our work, rather than concentrating on the data post-processing, we emphasize on the baked data itself. Moreover, our scenario involves multi-NeRF rendering, where maximizing the rendering quality of a single NeRF network is no longer practical, as the overall cost could easily exceed the device's capabilities. Instead, it is essential to consider the combined rendering quality and cost of multi-NeRFs of all the segmented objects to ensure a satisfactory user experience. These factors significantly increase the complexity of our configuration optimization.

\subsubsection{Problem formulation and analysis}
Let the size of the NeRF representation for each object be represented as $s_i$ and the corresponding rendering quality as $q_i$. Given a fixed memory size $H$, our goal is to find the configuration pair $(\theta_i = (g_i, p_i))$ for each object $i$'s NeRF representation that maximizes the total quality $\sum q_i$, while ensuring that $\sum s_i \leq H$. As $s_i$ and $q_i$ can be represented by $f_{si}(\theta_i)$ and $f_{qi}(\theta_i)$. The configuration selection problem can be formulated as follows:
\begin{equation}
\begin{aligned}
    maximize&\sum^n_{i=1}\sum_{\theta_{j}\in C_i}f_{qi}({\theta_{j}})x_{ij} \\
    subject\ to&\sum^n_{i=1}\sum_{\theta_{j}\in C_i}f_{si}({\theta_{j}})x_{ij} \leq H \\
    & \sum_{\theta_{j}\in C_i}x_{ij} = 1,\ i=1,2,3,..,n  \\
    & x_{ij}\in\{0, 1\}, i = 1,2,3,...,n
\end{aligned}
\label{opti_equ}
\end{equation}
where $C_i$ denotes the configuration space for object $i$, while $x_{ij}$ indicates whether $\theta_j$ is selected as the configuration for object $i$. After careful analysis, we reveal that the multi-multip configuration selection problem essentially can be interpreted as a multiple-choice knapsack (MCK) problem \cite{kellerer2004multiple}. Specifically, the memory constraint $H$ acts as the knapsack's capacity, and the goal is to select the optimal configurations for all the objects that need to be rendered among the corresponding configuration space $C_i$ to maximize the total rendering quality. Since the MCK problem is NP-hard \cite{kellerer2004multiple}, our multi-NeRF configuration selection problem is also NP-hard.

\subsubsection{Algorithm design}
Most of the common algorithms to solve the MCK problem are built on the basis of greedy search\cite{zemel1980linear}\cite{dyer1995hybrid}\cite{sinha1979multiple}\cite{dudzinski1987exact}. However, these algorithms often rely on the fulfillment of a crucial condition that for every configuration pair $\theta_j \in C_i$ satisfies the equation \ref{MCK_cond}.
\begin{equation}
    f_{si}(\theta_{j}) + \sum_{h = 1, ..., m, h\neq i}\min_{\theta_{t} \in C_h}f_{sh}(\theta_{t})\leq H
\label{MCK_cond}
\end{equation}
This condition ensures the existence of a feasible solution to the MCK-based optimization problem. In our scenario, for each object, it's unable for us to initially determine which configuration pairs can satisfy this object, especially since there are two control knobs. To effectively filter out the configuration pairs that do not meet the equation \ref{MCK_cond} and prevent sub-optimal solutions, we propose a dynamic programming (DP) algorithm to solve our configuration selection problem in pseudo-polynomial time. The detailed algorithm is presented in Algorithm\ref{Algo},

\begin{algorithm}[t]
  \caption{Dynamic-programming-based configuration selection}
  \label{Algo}
  \begin{algorithmic}[1]
  \Statex \textbf{Input:} 
  The set of segmented objects: \textit{I};
  Prediction models $f_{si}$ and $f_{qi}$ for each object $i\in I$;
Configuration space $C_i$ for each object $i\in I$;
The size limitation $H$
  \Statex \textbf{Output:}
 The configuration list: $config$
  \State $\theta^*_i \gets min(C_i) = (min(g_i), min(p_i))$
  \State Initialize an empty array $q$
  \State Initialize an empty two-dimensional array: $choices$
  \State Initialize an empty list: $configs$
  \For{each object $i$ in $I$}
    \For{$j$ from $H$ down to 0}
        \For{$\theta$ in $C_i$}
            \State $r_i \gets H - \sum_{h = 1, ..., m, h\neq i}\min_{\theta_{t} \in C_h}f_{sh}(\theta_{t})$ 
            \If{$f_{si}(\theta) \geq r_i$}
                \State continue
            \EndIf
            \If{$j \geq f_{si}(\theta)$}
                \If{$q[j] < q[j - f_{si}(\theta)] + f_{qi}(\theta)$}
                    \State $q[j]\gets q[j - f_{si}(\theta)] + f_{qi}(\theta)$
                    \State $choices[i][j] \gets \theta$
                \EndIf
            \EndIf
        \EndFor
    \EndFor
  \EndFor
  \For{$m$ from ($size(I) - 1$) down to 0}
    \If{$choices[m][H]$ is not none}
        \State Add $(m, choices[m][H])$ to $config$
    \EndIf
  \EndFor
  \State \textbf{return} $config$
  \end{algorithmic}
\end{algorithm}

We first initialize two key arrays: the array $q$, which is used to store the highest quality that can be achieved under each size capacity under the limit, and the array $choices$, which is used to store the chosen configuration pair for the objects under the corresponding size capacity. Through nested loops iterating over object categories $I$, size capacity, and the configuration space $C_i$ for object $i$, the algorithm uses value $r_i$ to remove all the configuration pairs of object $i$ that can not satisfy the equation \ref{MCK_cond} and then progressively calculates the size and rendering quality of each object under specific configuration pair based on the generated profiler, using a state transition function to update the optimal quality values under the iterated corresponding size. Specifically, for each size capacity $j$ lower than $H$, the algorithm optimizes the array $q$ by evaluating the selection of the current object, thereby facilitating state transition which is shown as line 12 to 14 in algorithm \ref{Algo}. Ultimately, the algorithm backtracks through the $choices$ array to output the maximized quality value along with all the objects' corresponding selected configuration pairs.

The algorithm\ref{Algo} has a complexity that can be broken down based on its key components. First, for each object in the input set, we calculate its threshold size $r$, which has a time complexity of $O(n)$. Then, the dynamic programming process for each object's configuration selection takes $O(hc)$, where $h$ is the value of the input size constraint, and $c$ refers to the size of the object's configuration space $C$. Therefore, the overall complexity for the algorithm is $O(nhc)$.

\section{Evaluation}
\subsection{Evaluation setup}
\textbf{Devices}: We use two commercial mobile devices with different capabilities, iPhone 13 and Pixel 4, to evaluate the performance. iPhone 13 is equipped with a CPU featuring cores at 3.23 GHz and 2.01 GHz, along with 4GB of memory. Pixel 4 is equipped with a CPU featuring cores at 2.84 GHz, 2.42 GHz, and 1.78 GHz, along with 6GB of memory. 
\\
\textbf{Dataset}: We conduct our evaluation experiments on simulated scenes that consist of different synthetic 360-degree objects as well as real-world scenes. These objects come from the dataset provided by \cite{mildenhall2021nerf} and real-world scenes come from the dataset provided by \cite{mildenhall2019local}. These datasets contain objects and scenes with corresponding image sets for training and testing.
\\
\textbf{Baselines}: We compare the overall performance of NeRFlex with both the single NeRF representation and Block-NeRF\cite{tancik2022block}, a typical multi-NeRF rendering method. We also evaluate the performance of our configuration selector against two others within the NeRFlex framework: the average-size-based (Fairness) method and the sequential least squares programming (SLSQP) method.
\begin{itemize}
    \item The single NeRF representation. The whole scene is rendered by a single NeRF. As previously mentioned, NeRF models designed for complex scene rendering on mobile devices are often simplified representations of high-precision and complex NeRFs. To demonstrate the performance of our rendering, we have also chosen the initial full-scale models used in MobileR2L\cite{cao2023real} and Re-rend\cite{rojas2023re}, namely NGP and Mip-NeRF 360. Another baseline is mobileNeRF, as it is widely recognized and applied in related work and offers high rendering quality on mobile devices. The configuration pair for this single nerf is $(g, p) = (128, 17)$ as recommended in the original paper\cite{chen2022mobilenerf}. 
    \item Block-NeRF. This NeRF variant is a typical multi-NeRF framework. For this rendering algorithm, each object in the scene is represented independently by a separate mobileNeRF. The configuration for each NeRF remains the same as the single NeRF, which is $(g, p) = (128, 17)$.
\end{itemize}

\begin{figure}[!tbp]
    \centering
    \subfigure[MobileNeRF (Memory: 201MB)]{\includegraphics[height=3.7cm,width=0.49\linewidth]{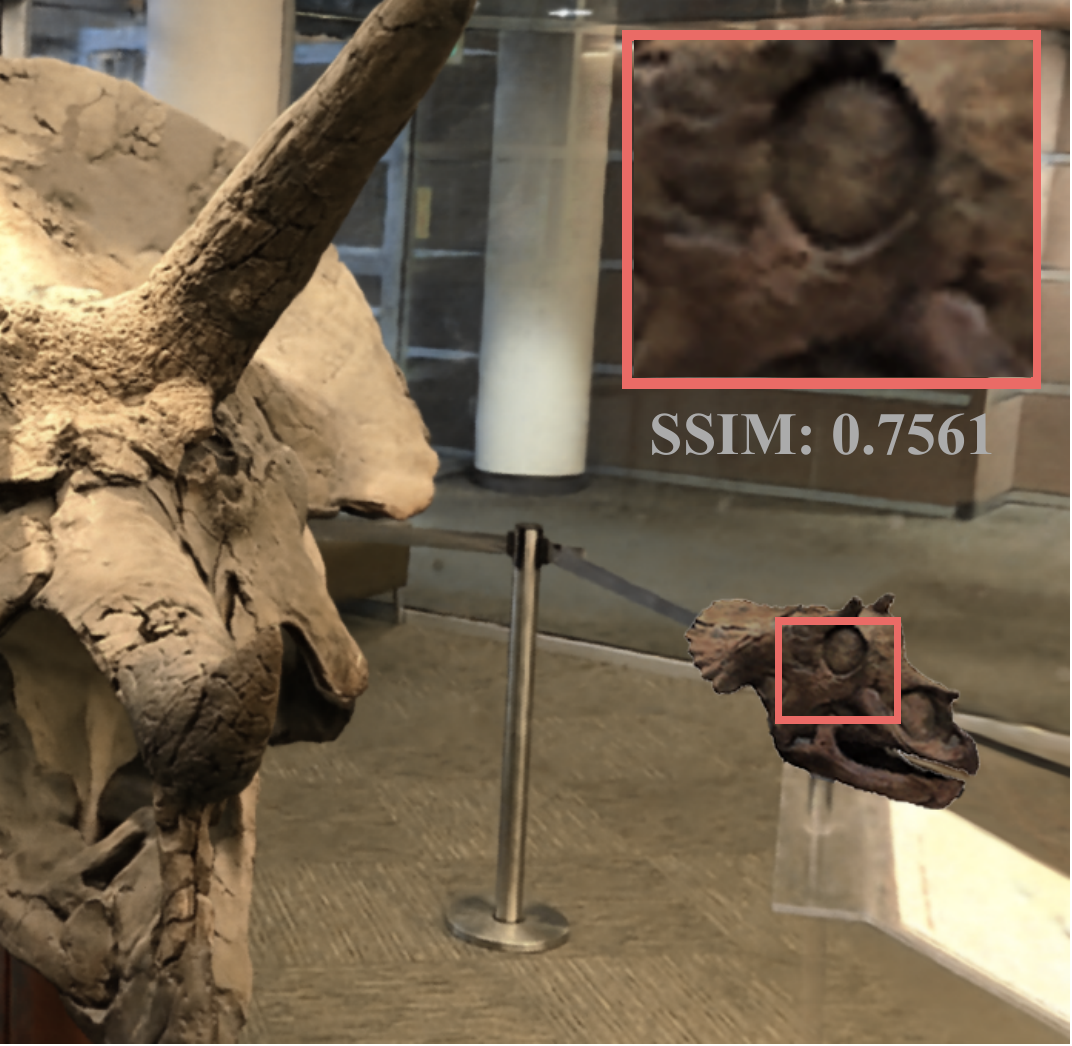}\label{visual_mobile}}
    \subfigure[MipNeRF 360]{\includegraphics[height=3.7cm,width=0.49\linewidth]{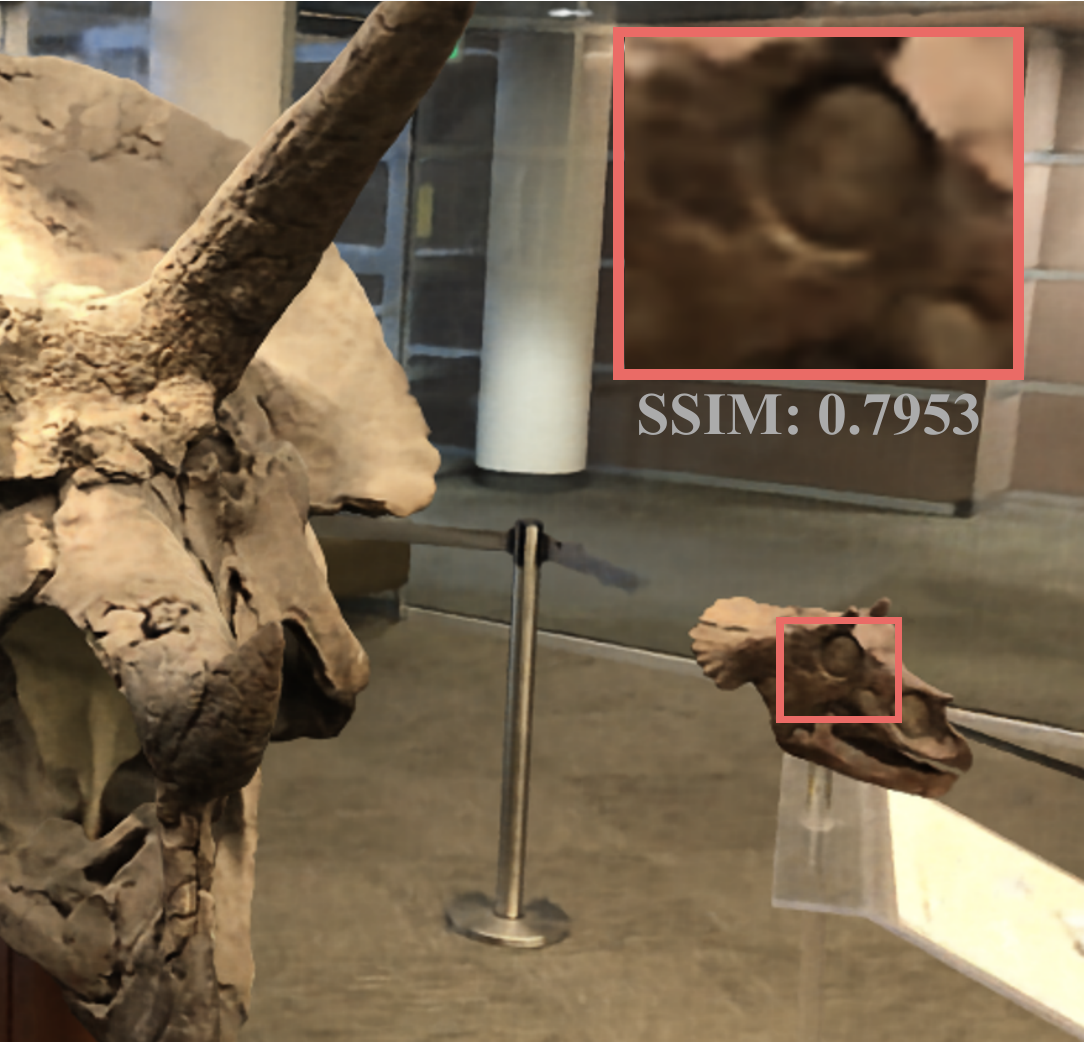}\label{visual_mip}}
    \subfigure[NGP]{\includegraphics[height=3.7cm,width=0.49\linewidth]{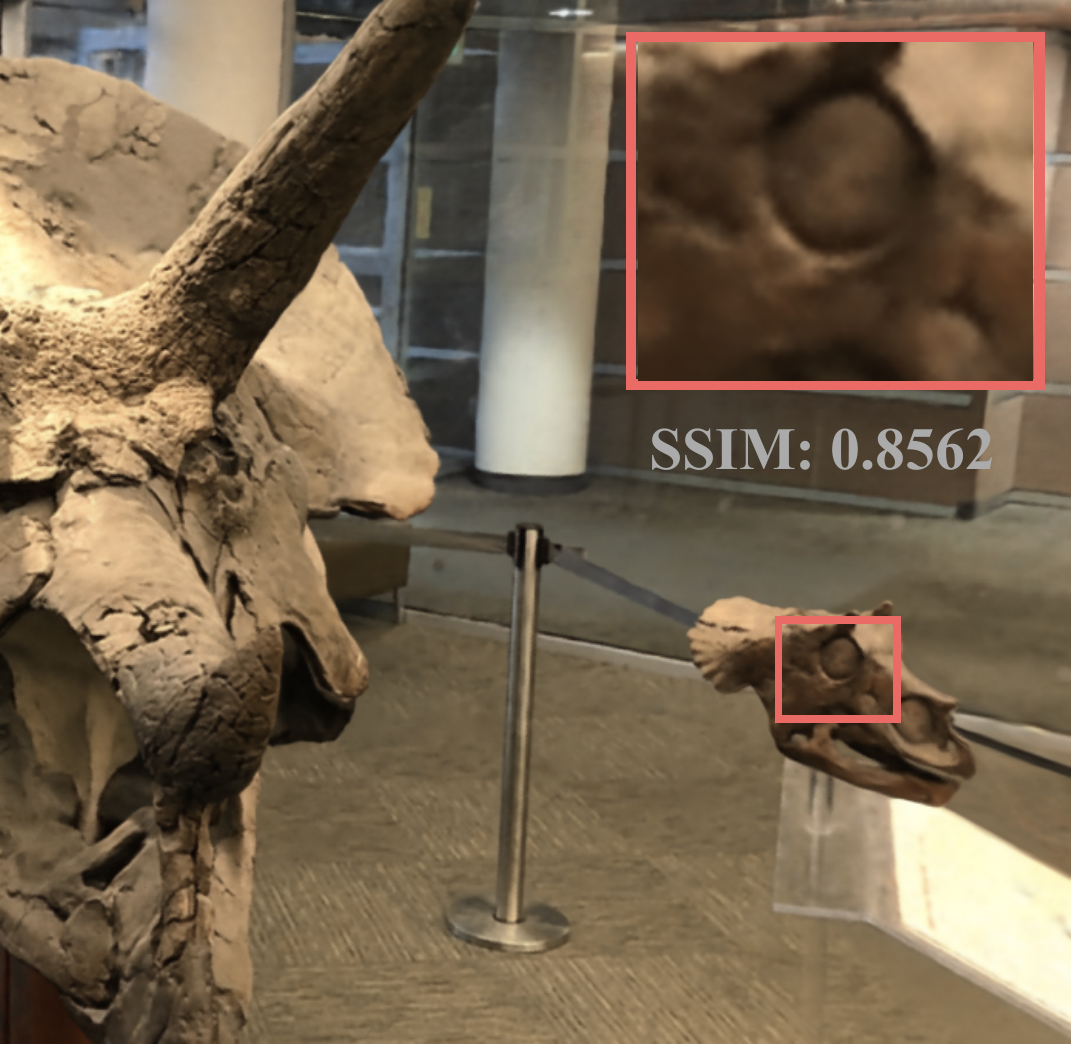}\label{visual_ngp}}
    \subfigure[Block-NeRF (Memory: 513MB)]{\includegraphics[height=3.7cm,width=0.49\linewidth]{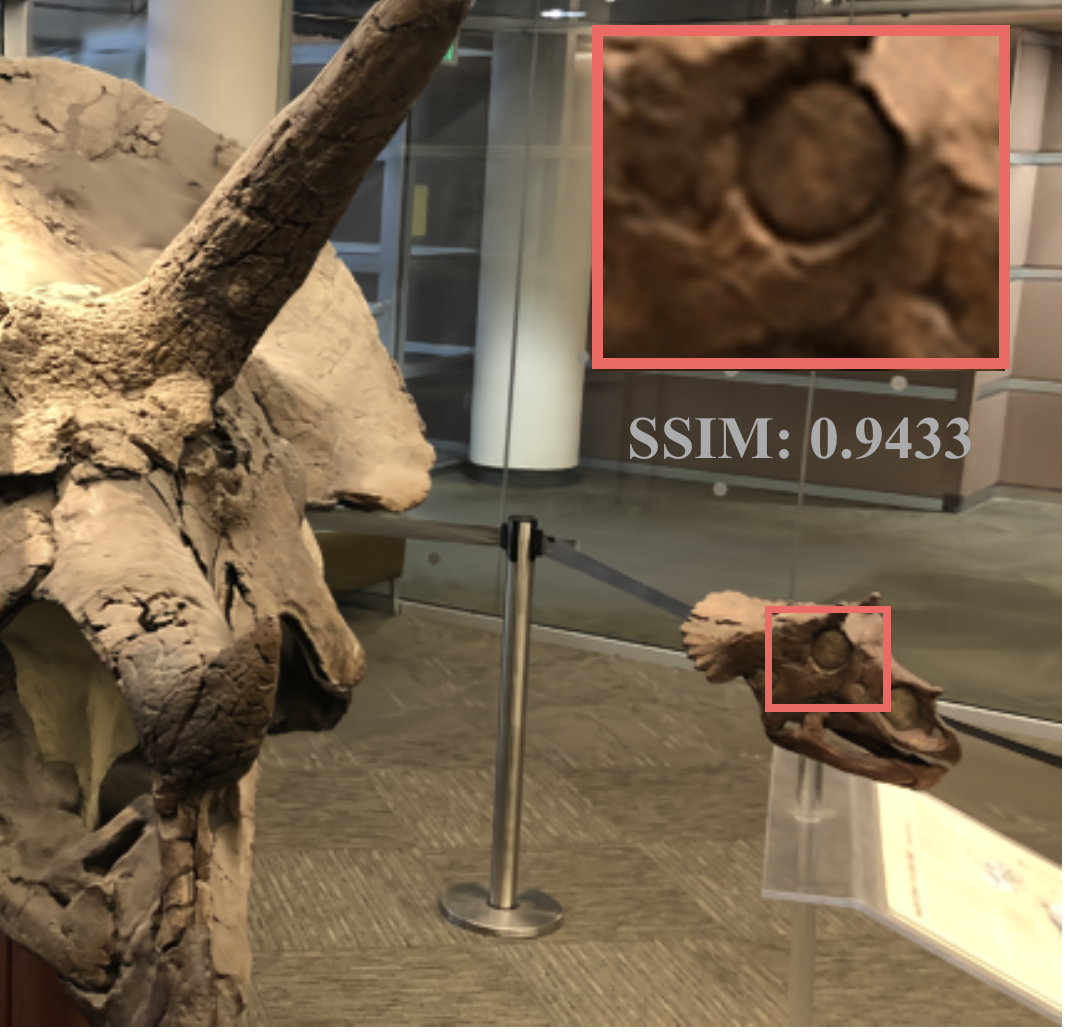}\label{visual_block}}
    \subfigure[NeRFlex (Memory: 240MB)]{\includegraphics[height=3.7cm,width=0.49\linewidth]{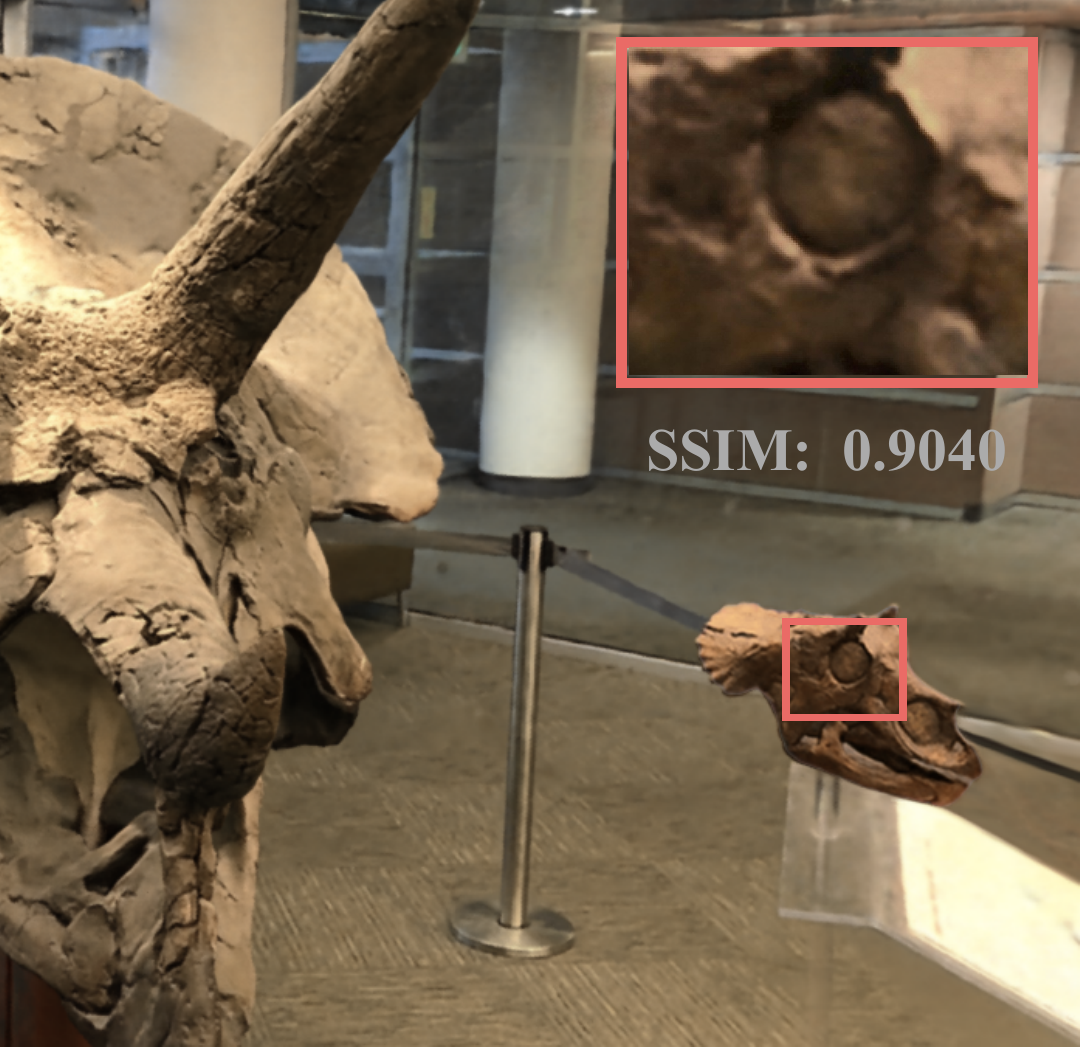}\label{visual_nerflex}}
    \subfigure[Ground Truth]
    {\includegraphics[height=3.7cm,width=0.49\linewidth]{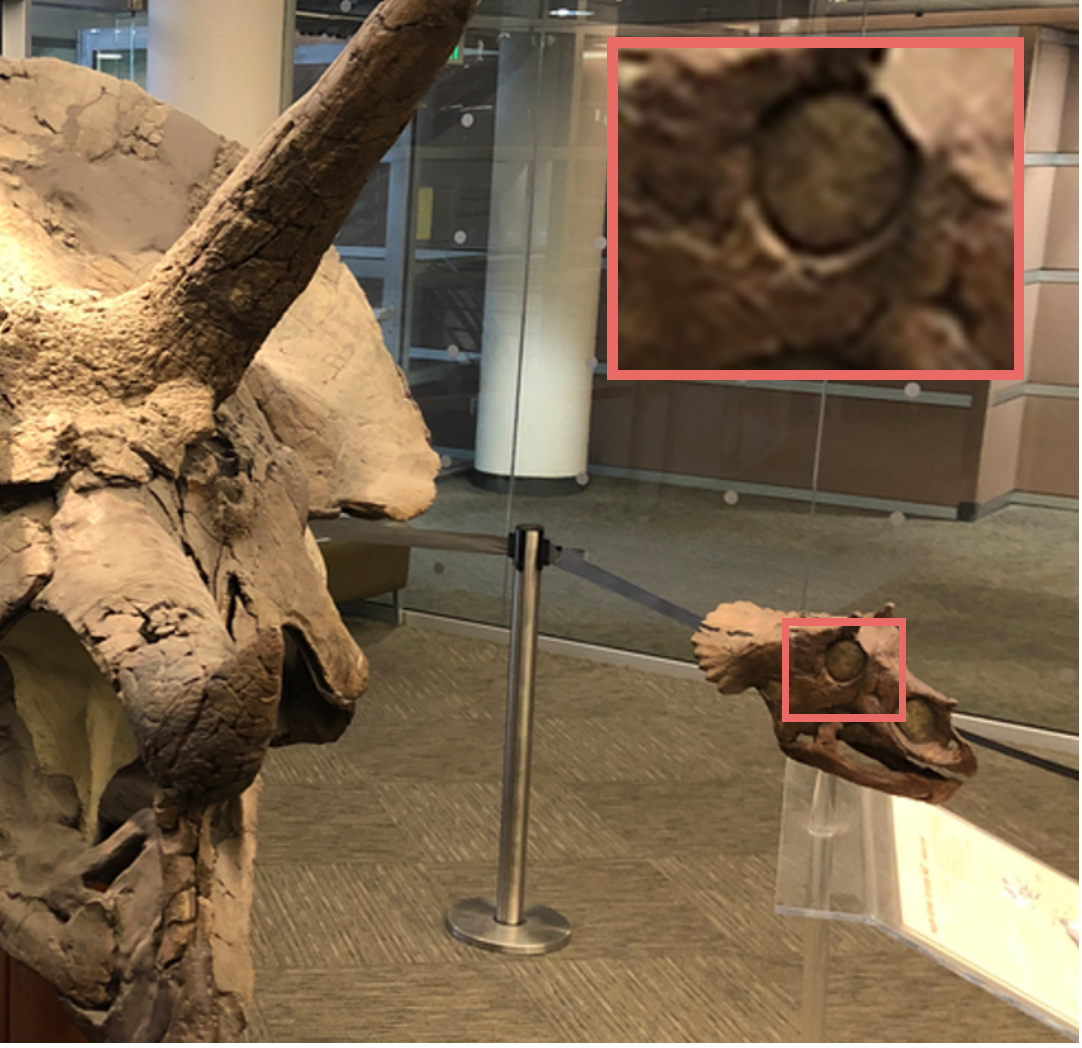}\label{visual_gt}}
    \caption{Rendering results of a complex scene with SSIM scores for high-frequency detail region on iPhone 13 (memory constraint: 240MB): NeRFlex outperforms baselines in visual quality and meeting the memory constraint. Single-NeRF model yields poor quality. Block-NeRF achieves the highest rendering quality but is not applicable in mobile settings.}
    \label{visual_results}
\end{figure}

\textbf{Metrics}: We evaluate the performance from three dimensions, which strongly affect user's viewing experience on their mobile devices.
\begin{itemize}
    \item Rendering visual quality: 
    We utilize three quantitative metrics to evaluate the rendering performance of NeRFlex. 
    
    \textbf{SSIM}: Structural Similarity Index Measure, which assesses the similarity between two images based on luminance, contrast, and structure. An SSIM value of 1, the maximum achievable score, signifies that the two images are identical \cite{wang2004image}. 
    
    \textbf{PSNR}: Peak Signal-to-Noise Ratio, which quantifies the difference between two images by calculating the mean squared error (MSE) of their pixel values. This provides a numerical assessment of the differences between the images \cite{hore2010image}. 
    
    \textbf{LPIPS}: Learned Perceptual Image Patch Similarity, which uses pre-trained deep neural networks to extract high-level semantic features from images. By computing the difference in feature space, LPIPS effectively measures perceptual similarity between images \cite{zhang2018unreasonable}.
    
    \item Data size: The generated multi-modal NeRF representation data size which can be used to reflect both the resource utilization and computation burden during the rendering. 
    \item Rendering smoothness: We use the frames per second, known as the FPS value to reflect the rendering fluency. Higher FPS generally means that users can enjoy a smoother experience when viewing NeRF rendering.
\end{itemize}

\textbf{Implementation}
In our work, we use WebGL as the rendering engine, as it is supported on nearly all mobile devices and integrated into their browsers.WebGL’s well-established optimizations for 3D rendering make it ideal for handling mesh grids \cite{rego20153dmol}. We use Safari for iPhone and Chrome for Pixel. For the segmentation module, we set the lowest maximum frequency among all the objects in the dataset as the threshold value. 
This setting allows us to allocate as many NeRF networks as possible, thereby maximizing rendering quality. From an evaluation perspective, this setting also enables us to test our system's performance on a larger problem with more decision variables.
During the evaluation, we write scripts to have the objects rotate at a fixed speed (7.5 seconds per 360 degrees) to ensure uniform viewing and that each pixel is re-rendered by NeRF for fair comparison. The entire evaluation was conducted on a cloud server equipped with a single Geforce RTX 3090 GPU, featuring 24GB of memory. The size limit for multi-modal NeRF representation data for iPhone 13 is set to 240MB and for Pixel 4, this value is set to 150MB. Based on our experiments, once the data size exceeds 240MB on iPhone, the local WebGL rendering engine fails to load the data. For the Pixel 4, when the data size exceeds 150MB, the average FPS drops by approximately 15, leading to a significantly degraded viewing experience with noticeable stuttering. Although this device has a higher memory capacity to handle larger files, its lower computational power significantly reduces viewing smoothness. Larger files include more data to process, further worsening performance.

\begin{figure}[!tbp]
    \centering
    \subfigure[Rendering quality performance.]{\includegraphics[height=3cm,width=0.48\linewidth]{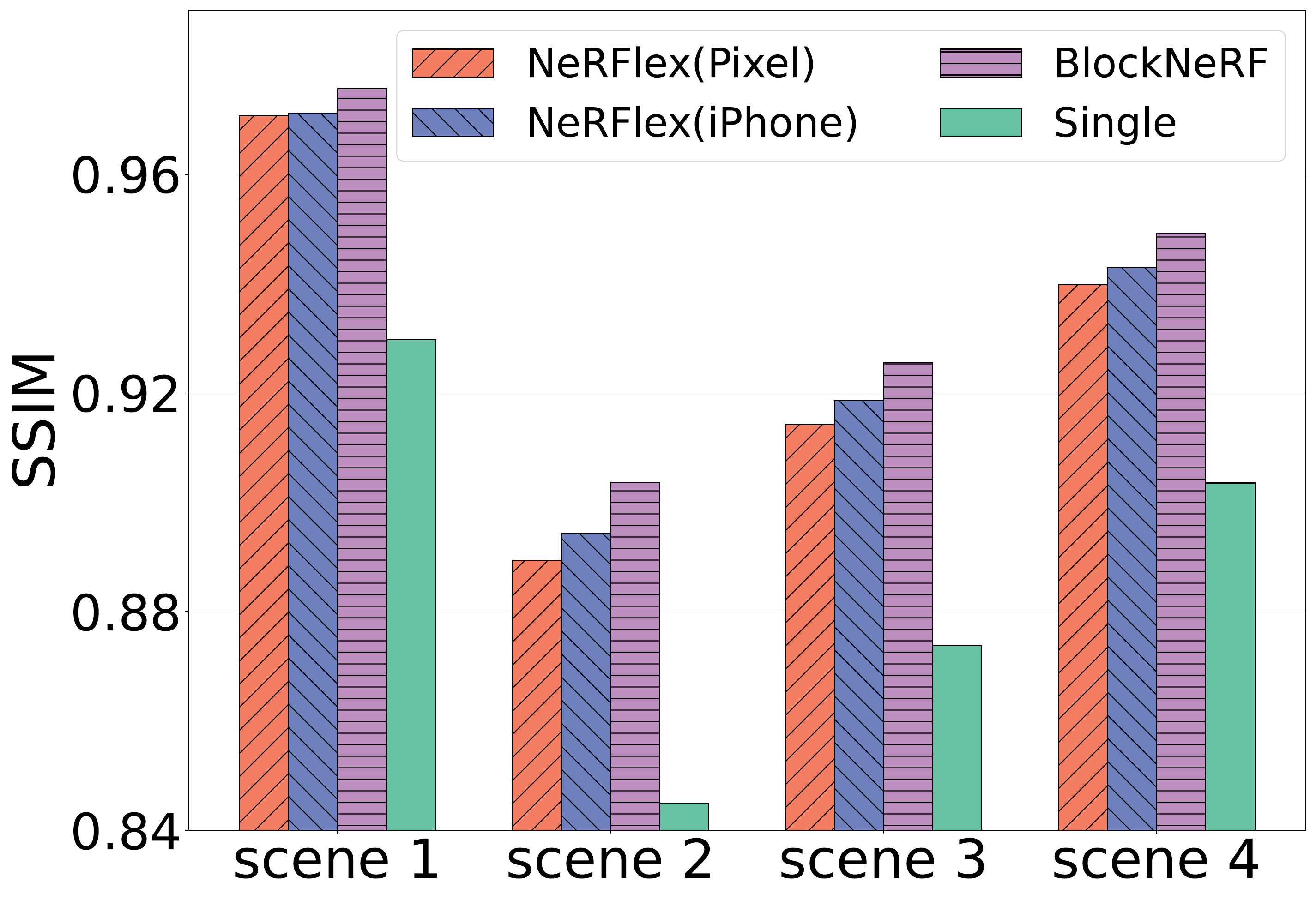} \label{quality_scene}}
    \subfigure[Size cost performance.]{\includegraphics[height=3cm,width=0.48\linewidth]{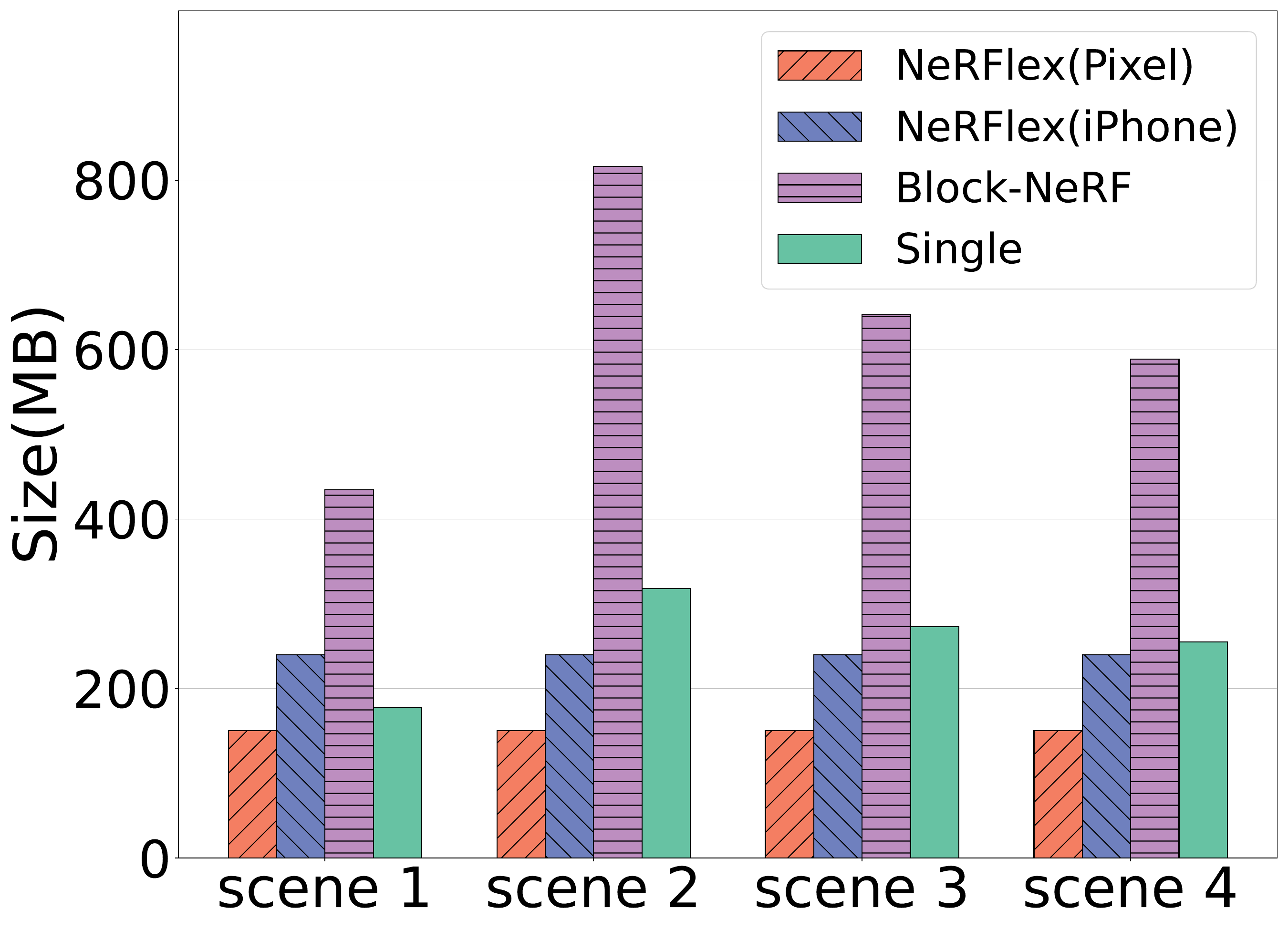}\label{size_scene}}
    \caption{The overall performance of different approaches on two representative mobile devices across different simulated scenes, the term Single refers to the MobileNeRF}
    \label{quality_sim_scene}
\end{figure}

\subsection{Overall Performance}

In this section, we first evaluate NeRFlex's overall performance, focusing on the metrics across different devices. Then, a more in-depth sensitivity analysis is conducted to reveal the reasons behind the superiority. 
For the simulated scenes, we construct four different scenes with different geometric complexities on two mobile devices. Each scene contains five objects from the dataset\cite{mildenhall2021nerf}. The geometric complexity is quantified by the number of mesh quad faces generated at a given mesh granularity, higher face counts indicate greater geometric complexity. 
Scene 1 is made of objects with the lowest geometric complexity. Scene 2 is made of objects with the highest geometric complexity.   
Scene 3 randomly selects five objects; Scene 4 includes five exclusively different objects in the dataset.

\begin{table}[!tbp]
\caption{Quantitative evaluation on rendering quality of NeRFlex compared to previous work on the real-world scenes.}
\centering
\begin{tabular}{|c|>{\centering\arraybackslash}p{1.45cm}|>{\centering\arraybackslash}p{1.45cm}|>{\centering\arraybackslash}p{1.45cm}|}
\hline
\textbf{Method}& \textbf{PSNR} $\uparrow$ & \textbf{SSIM} $\uparrow$  & \textbf{LPIPS} $\downarrow$ \\ \hline
MipNeRF 360\cite{rojas2023re}\cite{barron2022mip} & 26.549 & 0.815 & 0.183 \\ \hline
NGP\cite{cao2023real}\cite{muller2022instant} & 27.212 & 0.851 & 0.136 \\ \hline
MobileNeRF\cite{chen2022mobilenerf} & 26.027 & 0.785 & 0.207 \\ \hline
NeRFlex & 27.651 & 0.886 & 0.114 \\ \hline
\end{tabular}

\label{quality_table}
\end{table}

\textbf{Rendering quality performance.}
In this section, we evaluate the rendering quality performance of our system across various scenes, including the simulated ones and the real-world ones.
The visual instances rendered by different baselines on real-world scene are presented in Figure~\ref{visual_results}. While MobileNeRF demonstrates notable detail reconstruction, the details may fail to align the ground truth accurately, and it exhibits significant errors in reproducing lighting effects along with artifacts in complex scene representation, leading to lower metric scores. In contrast, these instances highlight that NeRFlex delivers superior rendering quality in high-frequency detail regions compared to Single-NeRF-based representations, while achieving performance comparable to the Block-NeRF framework. Furthermore, unlike Block-NeRF, which requires memory resources far exceeding device capacity (540MB for the visualized test scene), NeRFlex efficiently restricts resource usage within the device's capabilities.
Figure \ref{quality_scene} demonstrates NeRFlex's rendering performance on different commercial mobile devices compared with the single NeRF and block NeRF framework across various simulated scenes.
Compared to the Single NeRF representation, the other two methods demonstrate significantly superior rendering quality under the multi-NeRF framework across all scenes. Notably, in Scene 2, which consists of complex objects, Single NeRF only achieves an SSIM of around 0.84, whereas the other methods exceed 0.88 SSIM on both devices. For scenes with lower complexity, NeRFlex achieves an SSIM that is 0.05 higher than Single NeRF on both devices. When compared to Block-NeRF, NeRFlex performs only 0.006 lower on average for iPhone. Even on the lower-performance device, Pixel, the average quality reduction is 0.01. 
The rendering quality quantitative comparison of NeRFlex across three rendering quality metrics, using real-world scenes and the baselines, is presented in Table \ref{quality_table}, further validating its performance. The results demonstrate that NeRFlex achieves better rendering quality on low-computation devices compared to various NeRF algorithms that rely on large computational resources for rendering complex scenes.

\textbf{Resource utilization performance.}
Figure \ref{size_scene} illustrates the resource utilization for different NeRF rendering methods on mobile devices across different simulated scenes. With the imposed size limitation, NeRFlex effectively limits the size of the generated NeRF representation data. In contrast, Block-NeRF’s minimum resource usage across all scenes exceeds 400 MB, far surpassing the capacity of the two devices used in our experiments. For complex scenes, such as Scene 2, it requires up to 800 MB. NeRFlex, however, only needs 150MB, one-fifth of that, to run on the Pixel. Moreover, Block-NeRF also consumes around 600 MB to render the other two scenes. While Single-NeRF significantly reduces resource demands compared to BlockNeRF, its data size still exceeds 250 MB for most scenes, making it impractical for devices. Additionally, its rendering quality is the lowest, as previously noted. This is because each image used to train Single-NeRF must fully contain the entire scene, which poses challenges in capturing high-frequency details along the trajectory. Despite using fine meshes to reconstruct these regions, Single-NeRF often fails to match real geometry accurately, sometimes introducing artifacts. Consequently, the increased model size does not translate into improved rendering quality, and the rendering quality may be even lower.

In summary, BlockNeRF achieves high rendering quality, but its resource demands are often too large to be rendered successfully on low-performance devices. While mesh-based Single-NeRF reduces the resource requirements compared to BlockNeRF, it still cannot ensure that the NeRF representation data for every scene remains within the device's capacity, lacking adaptability to different devices. Additionally, Single-NeRF delivers the lowest rendering quality among the methods. In contrast, NeRFlex maintains high rendering quality while significantly reducing the computational burden. It also adapts to different device performance levels, enabling successful scene rendering across a variety of devices.

\begin{figure}[!tbp]
    \centering
    \subfigure[The real-time FPS performance on iPhone]{\includegraphics[height=3.5cm, width=0.48\linewidth]{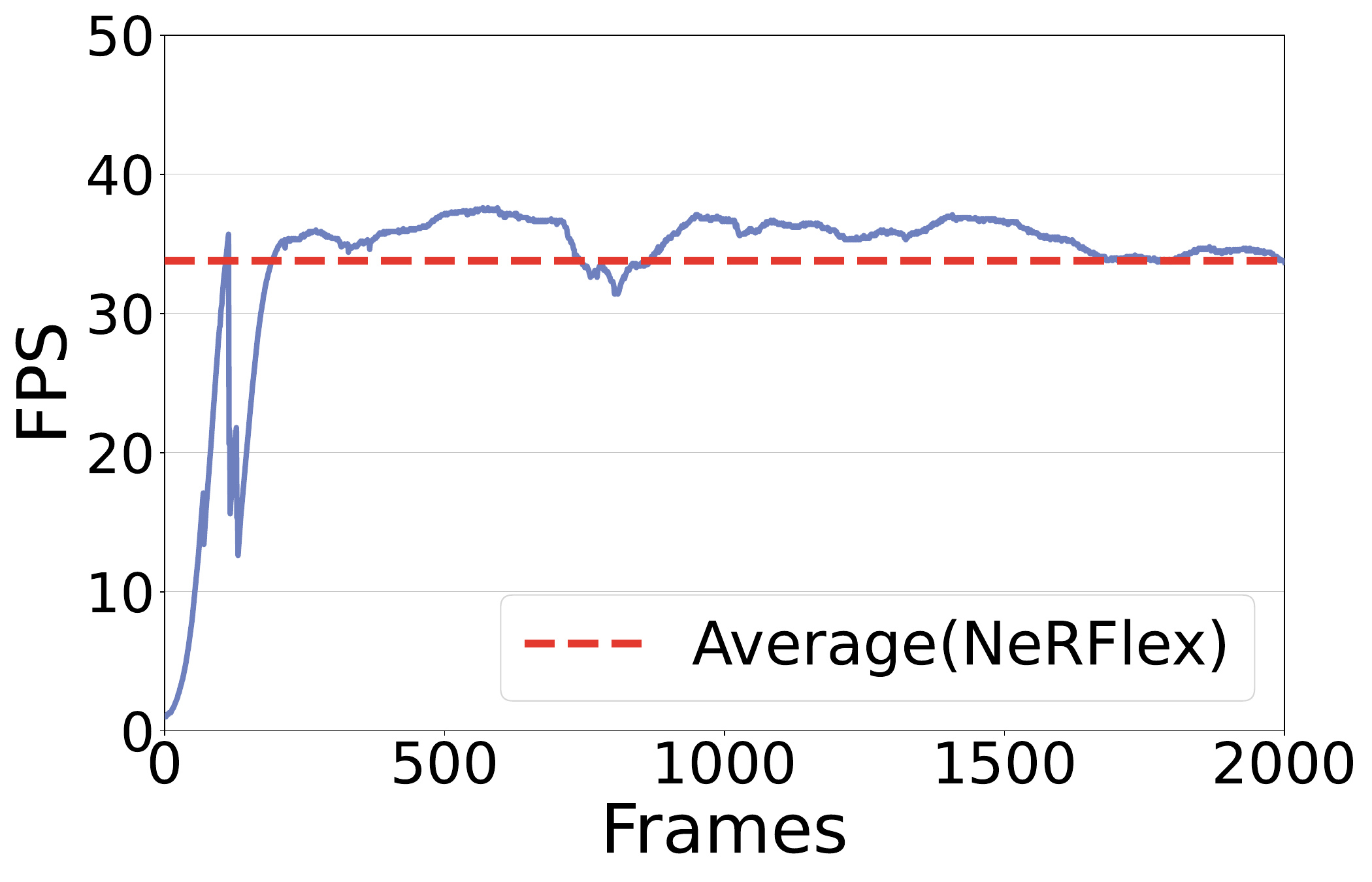} \label{iPhone_fps}}
    \subfigure[The real-time FPS performance on Pixel]{\includegraphics[height=3.5cm, width=0.48\linewidth]{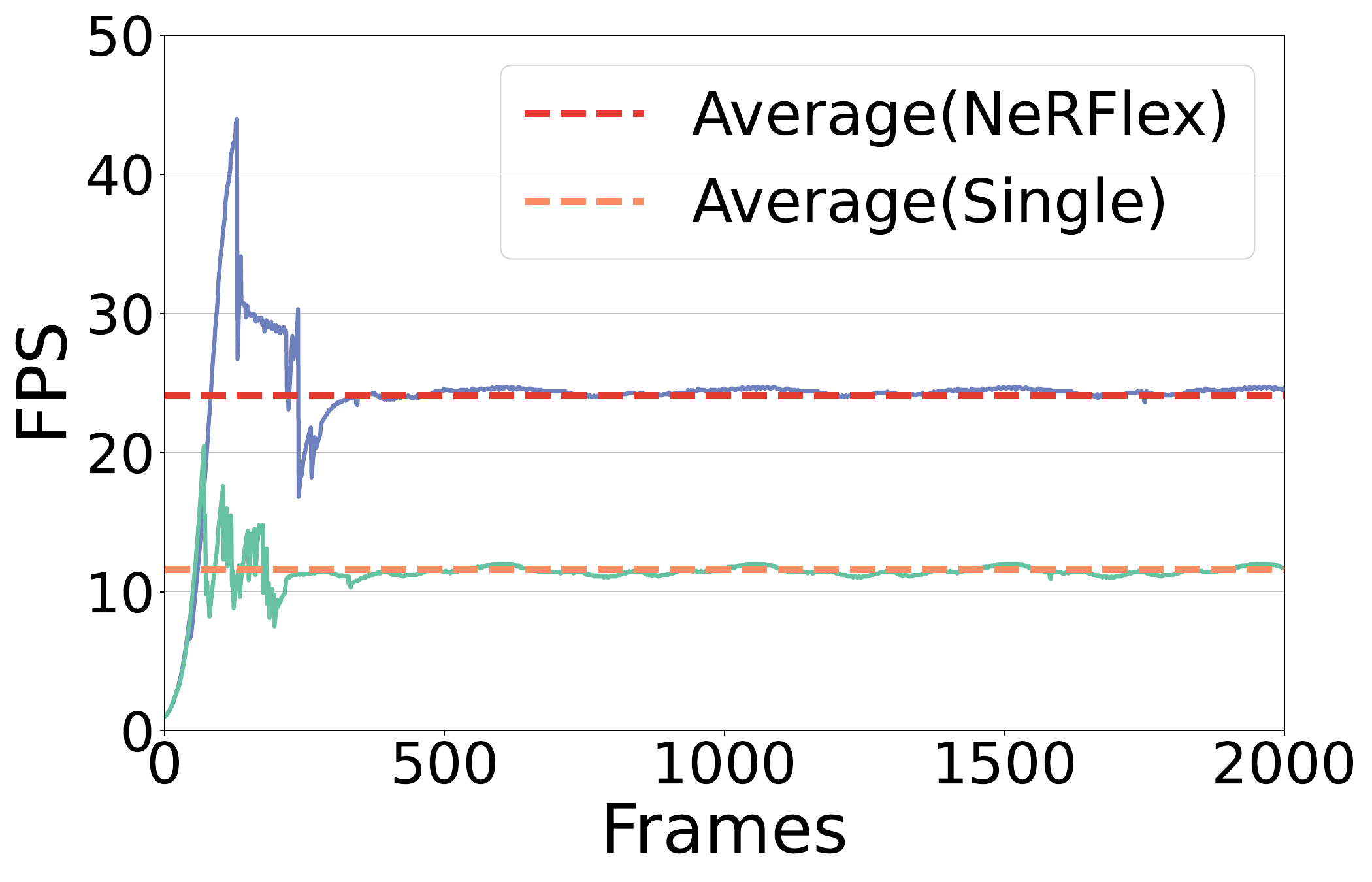}\label{Pixel_fps}}
    \caption{The FPS performance comparison between NeRFlex and Single NeRF across different devices. On the iPhone (left sub-figure), Single NeRF fails to render the scene as the memory requirements exceed the device's capacity, resulting in an FPS of 0. Similarly, Block-NeRF is unable to render any scenes on either device due to memory limitations.}
    \label{FPS_all_device}
\end{figure}

\textbf{Rendering smoothness performance.}
To better capture the randomness often present in the real world, we use Scene 3, a randomly generated scene, as the test case.
Figure \ref{iPhone_fps} and Figure \ref{Pixel_fps} illustrate the FPS performance on different mobile devices, namely iPhone 13 and Pixel 4, the term Single in these figures refers to MobileNeRF. 
The solid line represents the real-time FPS while the dash line reflects the average FPS during the rendering of 2000 frames. 
From the aspect of real-time performance, the FPS performance trends on these two devices have similar trends. Initially, the FPS rate exhibits significant fluctuations, which are primarily caused by
loading all the generated multi-modal NeRF files.
After that, the users should experience a relatively smoother viewing experience without sudden stuttering as shown in Figure \ref{FPS_all_device}. 
From the perspective of the overall viewing experience, thanks to the constraints we set based on the capabilities of each device, both devices were able to successfully render the scene with relatively high rendering smoothness.
According to Figure \ref{iPhone_fps}, the average FPS is around 35 during the viewing process. For the low-performance device, Pixel, shown in Figure \ref{Pixel_fps}, our system improves the FPS by 2 times compared to the single NeRF and can maintain around 25 FPS. The Block-NeRF requires extremely large computational resources, making it unable to complete rendering on either device.
Therefore, with the assistance of the designed modules, NeRFlex is able to provide a smooth rendering process for complex scenes on mobile devices, effectively addressing the issue of Block-NeRF and Single NeRF rendering in complex scenes, where the representation data often exceeds the device's memory capacity, leading to rendering failures.

\begin{figure}[!tbp]
    \centering
    \subfigure[Rendering quality performance on iPhone.]{\includegraphics[height=3.3cm,width=0.48\linewidth]{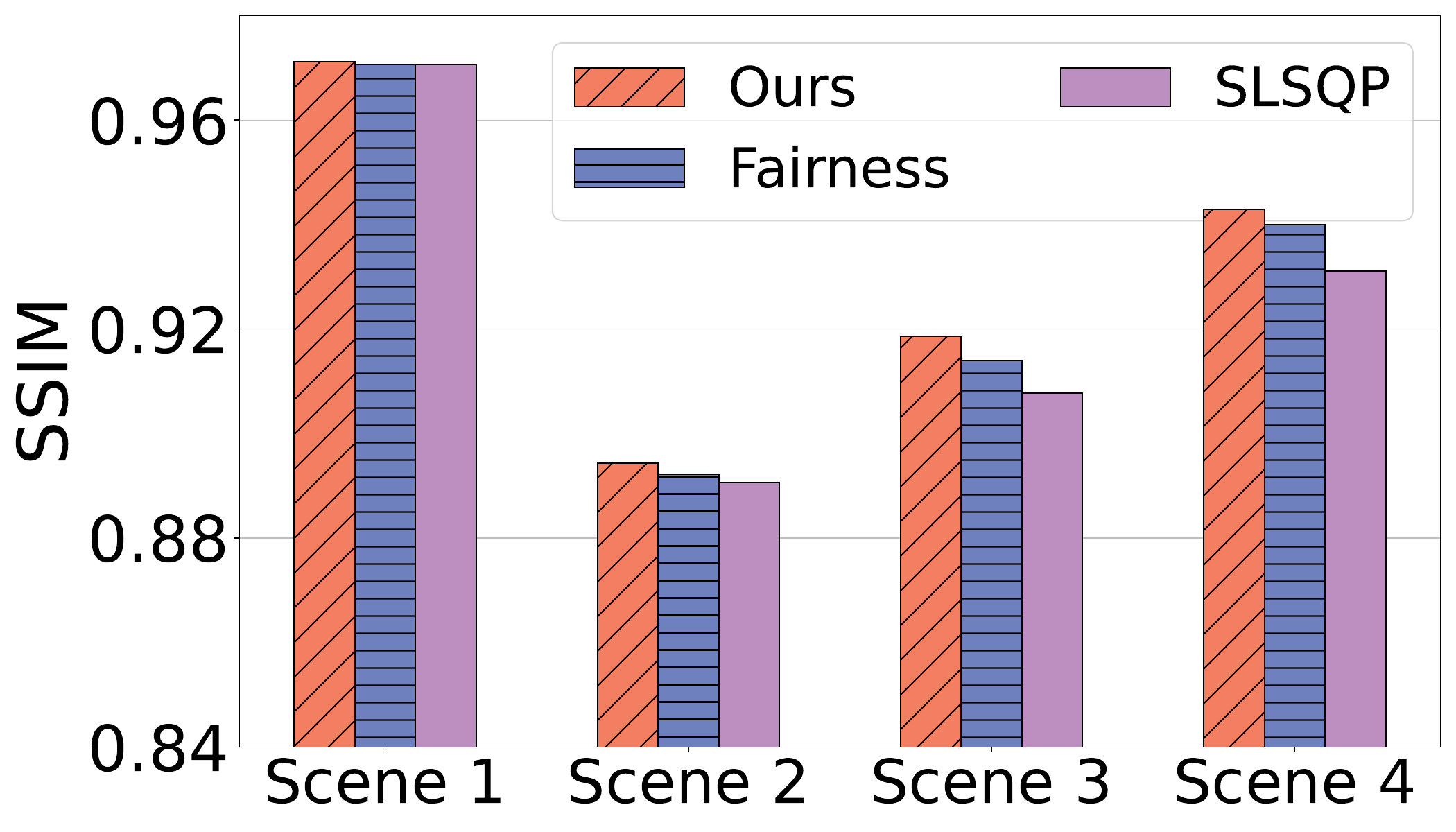} \label{sen_scene_iphone}}
    \subfigure[Rendering quality performance on Pixel.]{\includegraphics[height=3.3cm,width=0.48\linewidth]{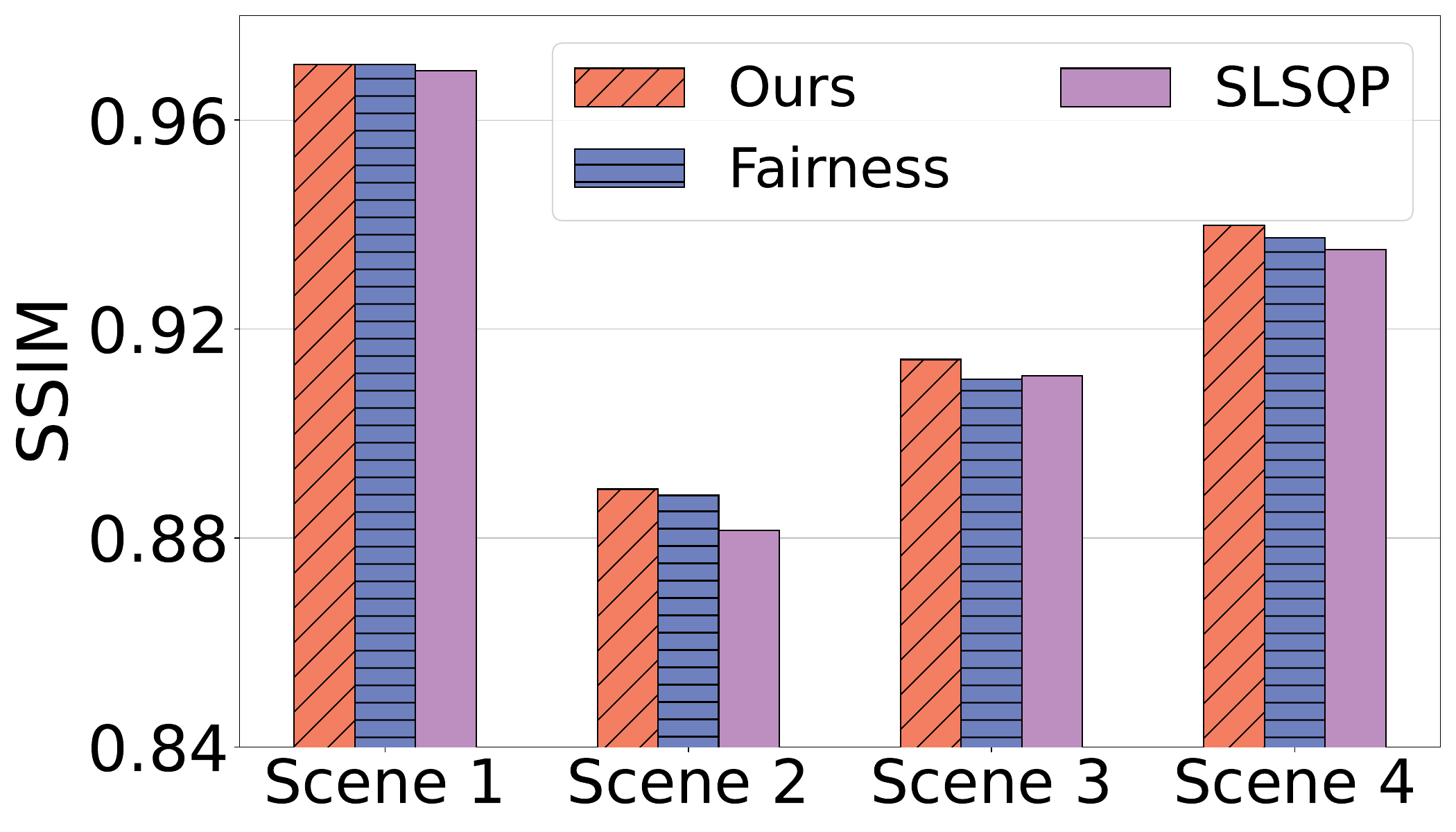}\label{sen_scene_pixel}}
    \caption{The quality performance of different approaches on two representative mobile devices}
    \label{sen_quality_scene}
\end{figure}

\subsection{Understanding NeRFlex's Performance}
In this section, we evaluate the performance of different configuration selection methods for the NeRFlex framework. We implement NeRFlex with Fairness: Rather than using our proposed DP algorithm to determine the configuration, this baseline divides the total size limit equally and allocates the same memory budget among the segmented objects. It then uses performance profilers to select the optimal configuration pair for each object, maximizing rendering quality within the allocated memory budget. We also implement NeRFlex with SLSQP: We use the SLSQP algorithm instead of our proposed algorithm to solve Eq.\ref{opti_equ}. The key concept of this algorithm is to approximate the gradient and Hessian matrix of the objective function using least squares, generating a search direction. It then solves a system of linear equations to update the optimization variables \cite{gill2011sequential}. The results are shown in the Figure \ref{sen_quality_scene}.

Among the three configuration selection methods under the NeRFlex framework, our method consistently outperforms the others. 
For the scenes composed of high-complexity objects, the performance of SLSQP lags behind the other two methods, especially on low-end devices. 
For scenes with high object diversity, consisting of both high and low-complexity objects, our method shows an obvious quality advantage, as demonstrated by the evaluations on Scene 3 and Scene 4 across both devices. On the iPhone, it achieves nearly 0.92 SSIM for the random scene, whereas the other two baselines only reach around 0.91. For Scene 4, we attain 0.94 SSIM on both devices, while the other methods struggle to achieve this level.

\textbf{High-quality rendering on complex objects.} 
To gain deeper insights into the reasons why NeRFlex consistently achieves higher quality, we conduct a more detailed analysis of Scene 4, as this scene has the highest object diversity.

\begin{figure}[!tbp]
    \centering
    \subfigure[The quality performance of each segmented object by different configuration selection methods. The left sub-figure presents the evaluation results on iPhone while the right sub-figure presents the evaluation results on Pixel.]{\includegraphics[height=3.5cm, width=\linewidth]{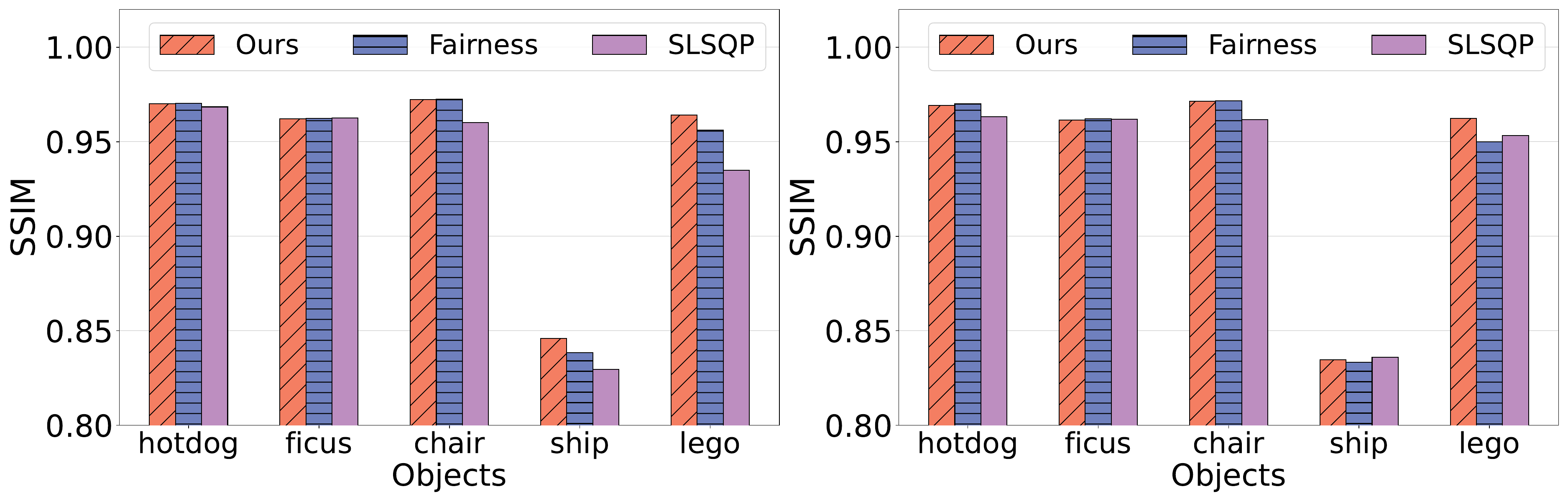}\label{quality_per}}\\
    \subfigure[The data size allocation details for each segmented object under different configuration selection methods on iPhone. ]{\includegraphics[width=\linewidth]{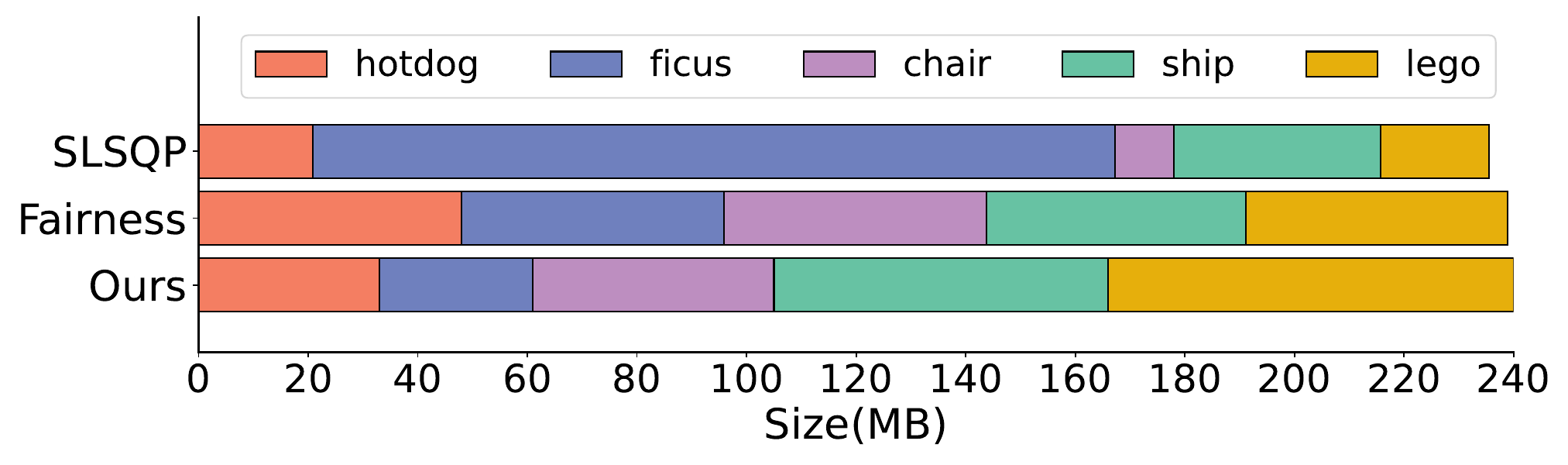}\label{size_pre_iphone}}
    \caption{The detailed performance for different configuration selectors on rendering quality and source allocation}
    \label{overall_com}
\end{figure}

Figure \ref{quality_per} illustrates the quality performance for different objects on the two devices. The orders of objects on the horizontal axis are arranged in 3D geometric complexity ascending order. From the evaluation results, our method always exhibits higher rendering quality for complex objects on different devices, which is especially evident on the high-capacity devices as shown in the left sub-figure of Figure \ref{quality_per}. In this figure, for the object ship, our method improves the quality by 0.01 and 0.02 compared with the Fairness and SLSQP respectively. For the Lego object, the improvement values are 0.01 and 0.03. For objects with lower geometric complexity, such as the hotdog, ficus, and chair, our method achieves comparable rendering quality to the baselines, with SSIM values consistently above 0.95 on both devices. 

\textbf{Optimized resource allocation.} Taking the Scene 4 rendering on the iPhone as an example, Figure \ref{size_pre_iphone} shows the detailed resource allocation for each object under different methods. The optimized resource allocation provided by our method is the key to its superiority. For low-complexity objects, our method allocates significantly fewer resources compared to the Fairness-based approach. For instance, the hotdog object only requires about 67\% of the resources allocated by the Fairness method, with minimal loss in rendering quality. This allows more resources to be allocated to objects that warrant improvement. The additional 14MB and 27MB allocated to the Lego and ship, respectively, result in significant quality improvements. On the other hand, although the SLSQP method also applies different allocation strategies for various objects, in this case, it produces an unreasonable resource allocation scheme. This issue likely arises due to two main factors: first, the choice of initial assumption values, as poor initial values can lead to incorrect optimization directions; second, the approximation error inherent in the SLSQP method may also contribute to this outcome.

In summary, our approach can intelligently allocate appropriate resources to different objects in the scene based on the capacity of various mobile devices, ensuring the best possible viewing experience of the overall scene for users.

\subsection{Overhead Analysis}
\begin{figure}[!tbp]
  \centering
  \includegraphics[width=\linewidth]{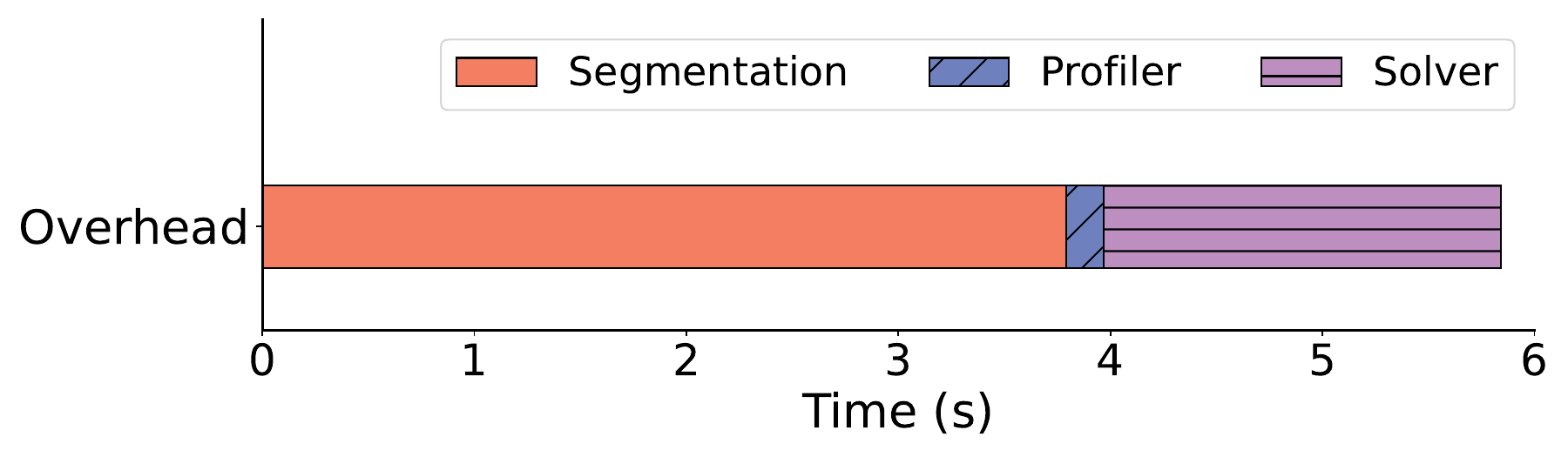}
  \caption{Execution time analysis for NeRFlex.}
  \label{overhead}
\end{figure}

In this section, we evaluate the overhead cost of our system, as detailed in Figure \ref{overhead}. NeRFlex can be generally divided into the cloud part for multi-NeRF preparation and the on-device rendering part.  We report the total processing time of different modules for processing twenty training images. 
In the multi-NeRF training and preparation stage, the image segmentation module takes approximately 3.8 seconds, which includes object recognition, segmentation, and interpolation, accounting for 64\% of the total processing time. This is primarily due to the need for operations like neural network-based semantic segmentation, frequency calculation, etc., across all training images. 
The performance profiler module requires 0.277 seconds, which is 4.7\% of the processing time. This includes the sample points generation and curve fitting. 
It takes 1.87 seconds, almost 31\% of the total time for our dynamic programming algorithm to find optimal configuration pairs. Overall, excluding neural network training, NeRFlex’s overhead cost for generating multi-modal NeRF representation data is about 5.9s. After this one-shot overhead, NeRFlex can achieve complex scene rendering with the multi-NeRF framework at around 25 FPS on mobile devices.

\section{Conclusion}
This paper presents NeRFlex, a NeRF-based rendering system designed for complex multi-object scenes on mobile devices, addressing two primary challenges: low rendering quality and high resource demands. To overcome these challenges, NeRFlex employs a multi-NeRF framework with detail-oriented segmentation and interpolation scaling, enabling the system to capture and represent fine details more effectively, thereby improving visual quality. Furthermore, the system adaptively configures the NeRF representation for each object, optimizing rendering quality within the limited computational resources of mobile devices. Experimental results demonstrate that NeRFlex successfully renders complex scenes across mobile devices with varying performance capacities, delivering superior smoothness and quality.

\section*{Acknowledgment}

This work is supported by the National Natural Science Foundation of China (Grant No. 62302292).

\bibliographystyle{IEEEtran}
\bibliography{sample}

\end{document}